\documentclass[preprint,12pt,5p]{elsarticle}

\usepackage{amsmath,amsfonts,amssymb,enumerate,amsthm}
\usepackage{relsize}
\usepackage{subfigure}
\usepackage{dsfont}
\DeclareMathOperator*{\argmax}{argmax}
\DeclareMathOperator*{\argmin}{argmin}

\DeclareMathSizes{10}{10}{10}{10}
\usepackage{bigints}
\usepackage{soul}

\usepackage[normalem]{ulem}

\newtheorem{thm}{Theorem}

\def\z{\mathbf{z}}
\def\p{\mathbf{p}}
\def\htau{\hat{\tau}}
\def\Sigma{{\sigma^{2}}}
\def\pfa{\bar{P}_{\text{FA}}}
\def\pfai{{P}_{\text{FA}}^{i}}
\def\pd{\bar{P}_{\text{D}}}
\def\pdi{{P}_{\text{D}}^{i}}
\def\Tx{L_{i}\left(\mathbf{r}\right)}
\def\Txx{T_{i}\left(\mathbf{r}\right)}
\def\Txa{L_{i}\left(\mathbf{r}\right)}

\def\brho{\frac{1}{\Sigma}}
\def\HTau{{\hat{\tau}_{i}}}
\def\Tau{\tau_{i}}
\def\Tau1{{\hat{\tau}_{i}}}
\def \X {X(d_{i})}
\def\T{\mathbf{T}}
\def\h{\mathbf{h}}
\def\x{\mathbf{x}}
\def\d{d_{i}}
\def\f{\mathbf{F}_{\mathbf{z}}}
\def\fz{F_{z}}

\def\lz{l(\HTau|\z)}

\def\CRLBAWGN{\text{CRLB}_{\text{1-D}}^{\text{AWGN}}}
\def\CRLBAWGN2{\text{CRLB}_{\text{2-D}}^{\text{AWGN}}}
\def\CRLBr{\text{CRLB}}

\def\df{f_{\hat{\tau}_{i}}(\hat{\tau}_{i}|\z)}

\def\dh{f_{|h_{i}|^{2}}(x)}
\def\MCRLB{\text{MCRLB}}
\def\hz{\hat{\mathbf{z}}}
\def\gammaprime{\gamma_{i}^{\prime}}

\def\absa{|h_{i}|}
\def\absaa{|h_{i}|^{2}}

\def\Bpfa{\bar{P}_{\text{FA}}^{\text{T}}}
\def\Bpd{\bar{P}_{\text{D}}^{\text{T}}}

\def\ti{\theta_{i}}
\def\dt{f\left(\theta_{i}\right)}

\def\HTaua{\hat{\tau}_{i1}}
\def\HTaub{\hat{\tau}_{i2}}

\def\r{\mathbf{r}}

\def\ri{r_{i}}
\def\rin{r_{i}[n]}
\def\htau{\hat{\tau}}
\def\Sigmai{\sigma_{i}}
\def\X{X(d_{i})}
\def\CRLBr{\text{CRLB}}

 \usepackage{graphics}

 \usepackage{graphicx}

 \usepackage{epsfig}

\usepackage{amssymb}

\journal{Physical Communication}
\date{}
\begin{document}

\begin{frontmatter}

\title{Location Estimation and Detection in Wireless Sensor Networks \\ in the Presence of Fading}
\author[label1]{Xue Zhang}
\author[label1]{Cihan Tepedelenlio\u{g}lu}
\author[label2]{Mahesh K. Banavar}
\author[label1]{Andreas Spanias} 
\author[label1]{Gowtham Muniraju}

\address[label1]{Sensip Center, School of ECEE, Arizona State University}
\address[label2]{Clarkson Center for Complex System Science ($C^{3} S^{2}$), Department of ECE, Clarkson University}
\date{}

\begin{abstract} 
In this paper, localization using narrowband communication signals are considered in the presence of fading channels with time of arrival measurements. When narrowband signals are used for localization, due to existing hardware constraints, fading channels play a crucial role in localization accuracy. In a location estimation formulation, the Cramer-Rao lower bound for localization error is derived under different assumptions on fading coefficients. For the same level of localization accuracy, the loss in performance due to Rayleigh fading with known phase is shown to be about $5$dB compared to the case with no fading. Unknown phase causes an additional $1$dB loss. The maximum likelihood estimators are also derived. 

In an alternative distributed detection formulation, each anchor receives a noisy signal from a node with known location if the node is active. Each anchor makes a decision as to whether the node is active or not and transmits a bit to a fusion center once a decision is made. The fusion center combines all the decisions and uses a design parameter to make the final decision. We derive optimal thresholds and calculate the  probabilities of false alarm and detection under different assumptions on the knowledge of channel information. Simulations corroborate our analytical results.

\end{abstract}


\end{frontmatter}

\section{Introduction}
\label{sec:introduction}
In many applications of wireless sensor networks (WSNs), the measured data are meaningful only when the location of the data is accurately known. The global positioning system (GPS) is widely used for outdoor localization applications \cite{patwari, ville, Han16, a2_unkerrors}. However, for indoor applications  \cite{Ballardini} and in non-LOS environments, GPS is not accurate. In such scenarios, wireless sensor networks (WSNs) \cite{sensor, Spanias}, which consist of low energy sensors, can be used for localization \cite{intro_positioning, Marc}. Ultra wide band (UWB) signals are often used for localization due to several advantages \cite{Poor}. In some applications, however, narrowband signals that are used for transmitting data have to be used for localization, as well. Therefore, localization using narrowband communication signals which are susceptible to fading needs to be studied. 

Localization in WSNs can be formulated as location estimation and location detection problems \cite{patwari, Mao, Neal03, RSS, ref3_Liang, ref1_Cota, ref2_Wang}. In the estimation formulation, the location of a node at an unknown location (target node) needs to be determined, with the help of nodes at known locations (anchors). Using noisy distance estimates obtained though transmissions between the anchors and the target node, the location of the target node is to be estimated. In contrast, in the detection formulation, the target node is in an known area; however, its exact position is unknown. The target node is not active all the time. When inactive, there is no transmission, and each anchor only receives noise. When the target node is active, each anchor receives signal (subject to fading) plus noise. In this case, a detection framework is needed to determine whether the target node is active or not. 

\begin{figure}[t]
	\centering
	\includegraphics[
	height=2in,
	]
	{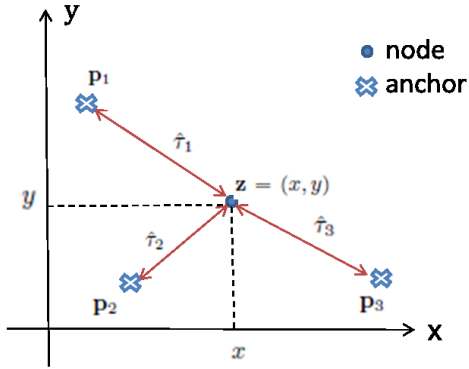}
	\caption{A wireless sensor network with $M=3$ anchors and $1$ node. The variable $\htau_{1}$ is the time of arrival measurement between the node located at $\z=(x,y)$ and the anchor located at $\p_{1}=(x_{1}, y_{1})$.}
	\label{fig:system model}
\end{figure}

Localization can be performed using range-based, direction-based, or hybrid methods \cite{Xue3, a6_rss_overview, a3_rlsnlos}. Typical range-based methods include received signal strength (RSS), time of arrival (TOA), and time delay of arrival (TDOA); the direction-based techniques include direction of arrival (DOA) \cite{FoutzDOA}; and hybrid methods \cite{ManikasLAA, a1_RSS_AoA}. In this paper, we select time of arrival (TOA) for both location estimation and detection problems, but the methodology can be applied to other measurement modalities as well. The accuracy of TOA measurements is highly dependent on signal bandwidth.
When the bandwidth is limited, the performance is affected by multipath fading and noise. Therefore, localization in the presence of fading needs to be studied.

In location estimation problems, the Cramer-Rao lower bound (CRLB) of the localization accuracy for TOA based and received signal strength (RSS) based approaches \cite{Xue3, patent}  have been studied in \cite{Neal03} in the absence of fading. Although some work has considered fading environments for TOA measurements \cite{Bergamo, fading}, derived the CRLB under the assumption that fading coefficients are deterministic unknown parameters \cite{Van1}, and considered a joint tracking and estimation framework \cite{a4_slat}, the CRLB of location estimation by considering fading coefficients as random unknown parameters has not been derived. 

In this paper, for the location estimation problems, the CRLB (Cramer-Rao lower bound) for localization error is derived under different assumptions on fading coefficients. These include cases where the fading coefficients are known at each anchor; unknown fading amplitude and phase with known distribution; and no CSI is available at any anchor. We show analytically that the loss in performance due to Rayleigh fading with known phase is about $5$dB compared to the case with no fading. Unknown phase causes an additional $1$dB loss.

In \cite{Neal03, a5_lbmse}, the maximum likelihood location estimator in the absence of fading has been derived. In our work, the maximum likelihood estimators under different fading scenarios are derived. These are compared with the estimator derived under the assumption of no fading, but deployed in a fading environment. 

Location detection in WSNs has been studied in \cite{Ray06}, which discretized the problem to obtain an $N$-ary hypothesis testing problem. In \cite{Van2}, a centralized sensor network with unknown fading coefficients has been considered. Although centralized methods may give a better performance, it is expensive in large WSNs. 
In \cite{xue_det}, a distributed location detection method in the absence of fading has been considered.  None of these works have studied localization under fading environments with explicit incorporation of the fading distribution. In \cite{RSSdetection}, a location verification system which uses RSS measurements for  detection under log normal shadowing is studied. It uses a centralized estimation and detection approach, whereas in this paper, each anchor makes its own decision on whether or not it detects an active node.

In this paper, a distributed location detection scheme is considered, where each anchor can make its own decision. Similar to the location estimation problem, different fading scenarios are considered. The probabilities of false alarm and detection are derived under different scenarios. 

The detection formulation is different from the estimation formulation in the following aspects. First of all, in detection problems, the goal is to detect the activity or silence of a node or multiple nodes at known locations; however, in estimation problems, the goal is to estimate the location of a node or multiple nodes, which are at unknown locations. Secondly, to estimate the location of a node, multiple anchors are needed in order to avoid ambiguity. For example, when using range-based methods, a minimum of two anchors are needed for one dimension (1-D), and three anchors are needed for two dimensions (2-D). On the other hand, to detect a node, each anchor can make a local decision on whether the node is active or not by correlating the received signal with the transmitted signal and then comparing with a threshold. The final decision can be made by exchanging this data with other anchors and a fusion center (FC). Therefore, the detection problem can be solved by using a distributed implementation based on exchange of bits between the anchors and a FC. Thirdly, the performance analysis is different for these two formulations. In the estimation formulation, the variance of the location estimation error is used as a performance metric, whereas for detection, metrics such as the probability of false alarm and the probability of detection are used \mbox{\cite{vantrees} \cite{Kay}}.

The rest of paper is organized as follows. In Section \ref{sec:estimation}, location estimation in the presence of fading is considered. The system model is proposed and three fading scenarios are considered. In Section \ref{sec:detection}, location detection in the presence of fading is studied. The probability of detection and probability of false alarm are derived under different fading scenarios. In Section \ref{sec:simulations}, the simulation results for both location estimation and detection are provided. Finally, in Section \ref{sec:conclusions}, concluding remarks are presented. 

\section{Location Estimation in the presence of fading}
\label{sec:estimation}
We assume a non-cooperative WSN, in which nodes do not communicate with each other. Further, we assume there are $M$ anchors and $1$ node in $\mathbb{R}^{n}$, where $n={1, 2}$. In 1-D, the location of the $i^{th}$ anchor, $\p_{i}=x_{i}$, and the node, $\z=x$ are scalars. In 2-D, $\p_{i}=[x_{i}, y_{i}]^{T}$ and $\z=[x, y]^{T}$ are vectors. Figure \ref{fig:system model} shows a sensor network with $M=3$ anchors and $1$ node. We assume the node communicates with all anchors. The measured TOA between the node and the anchor located at $\p_{i}$, is defined as $\htau_{i}$.
In location estimation, each anchor transmits a carrier modulated signal to a node, and the node transmits back immediately after it receives the signal. The two way transmission time is measured by each anchor, which can be halved to estimate the transmission time and distance. Define  $d_{i}=||\p_{i}-\z||_{2}$ as the true distance between the node located at $\z$ and the anchor located at $\p_{i}$. In the absence of fading, $\htau_{i}$ is Gaussian \cite{Helstrom}, 
\begin{equation}
\htau_{i} \sim \mathcal{N}\left(\frac{d_{i}}{c},\Sigma\right),
\label{Eq:Gassian}
\end{equation}
where $c$ is the speed of propagation of signals in the free space, and $\Sigma$ is the variance of the TOA measurements \cite{Neal03}. We will assume throughout that $\{\htau_{i}\}_{i=1}^{M}$ are independent.

\subsection{System Model}
\label{sec:estimation_system_model}
Define $h_{i}=|h_{i}|e^{j\ti}$ as the fading coefficient for the channel between the node and the $i^{th}$ anchor, where $\absa$ and $\ti$ are the amplitude and phase of the fading coefficient respectively, and $i \in \{1,2,\dots,M\}$. In the presence of fading, the statistics of $\htau_{i}$ is a function of $h_{i}$. In this paper, we consider the following scenarios: (a) $h_{i}$ is assumed to be known at each anchor; (b) $\ti$ is assumed to be known at each anchor, but $\absa$ is an unknown random variable with a known prior distribution; (c) No CSI (amplitude or phase) is available at any anchor. Although only 1-D and 2-D cases are considered, the results can be generalized to three dimension (3-D).

\begin{figure}[tb]
	\centering
	\centerline{\epsfig{figure=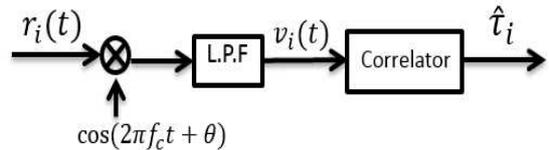,height=1in,width=3in}}
	\caption{Coherent TOA estimation scheme.}
	\label{fig:coherent}
\end{figure} 

\begin{figure*}[!t]
	
	\centering
	\subfigure[Non-coherent TOA estimation scheme.] 
	{
		\label{fig:noncoherent}
		\includegraphics[height=2in,width=3in]{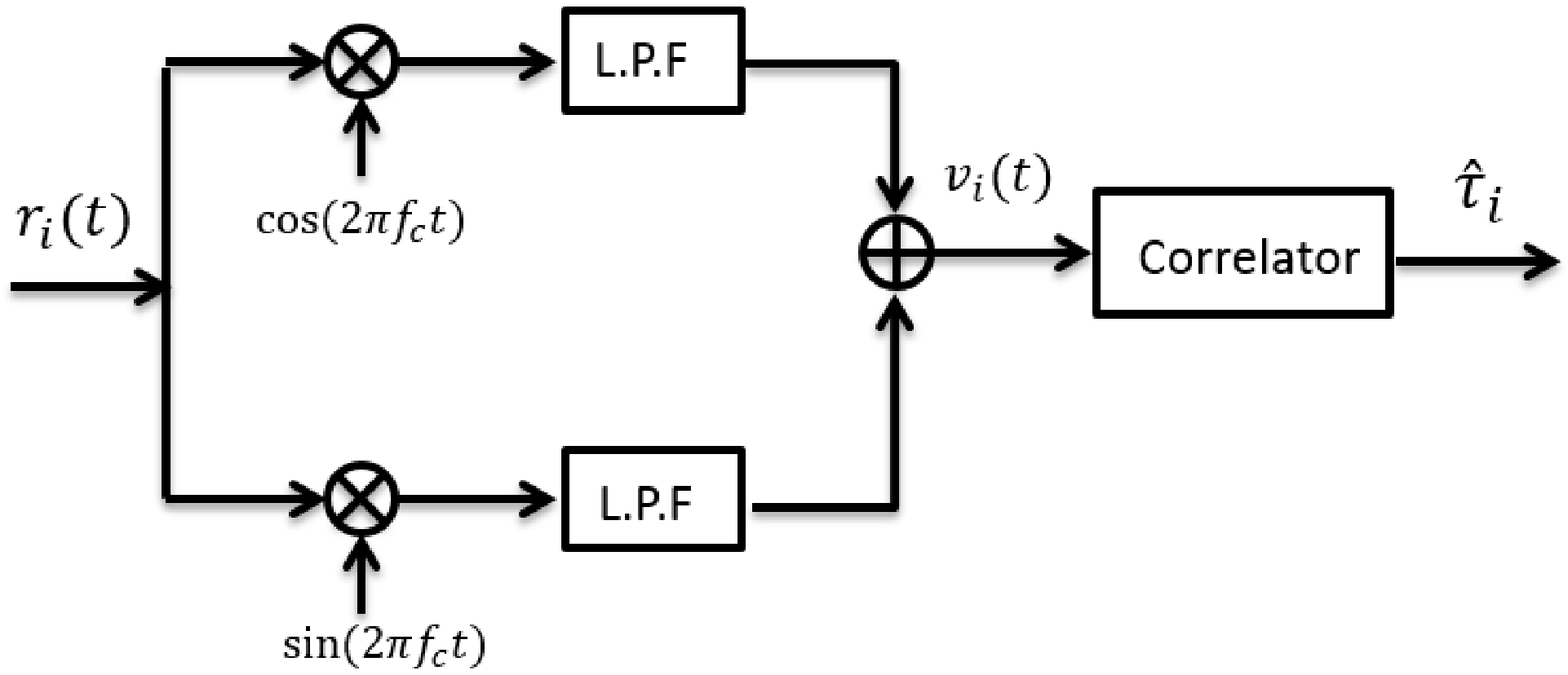}
	}
	\hspace{0.1cm}
	\subfigure[Alternate non-coherent TOA estimation scheme.] 
	{
		\label{fig:noncoherent2}
		\includegraphics[height=2in,width=2.8in]{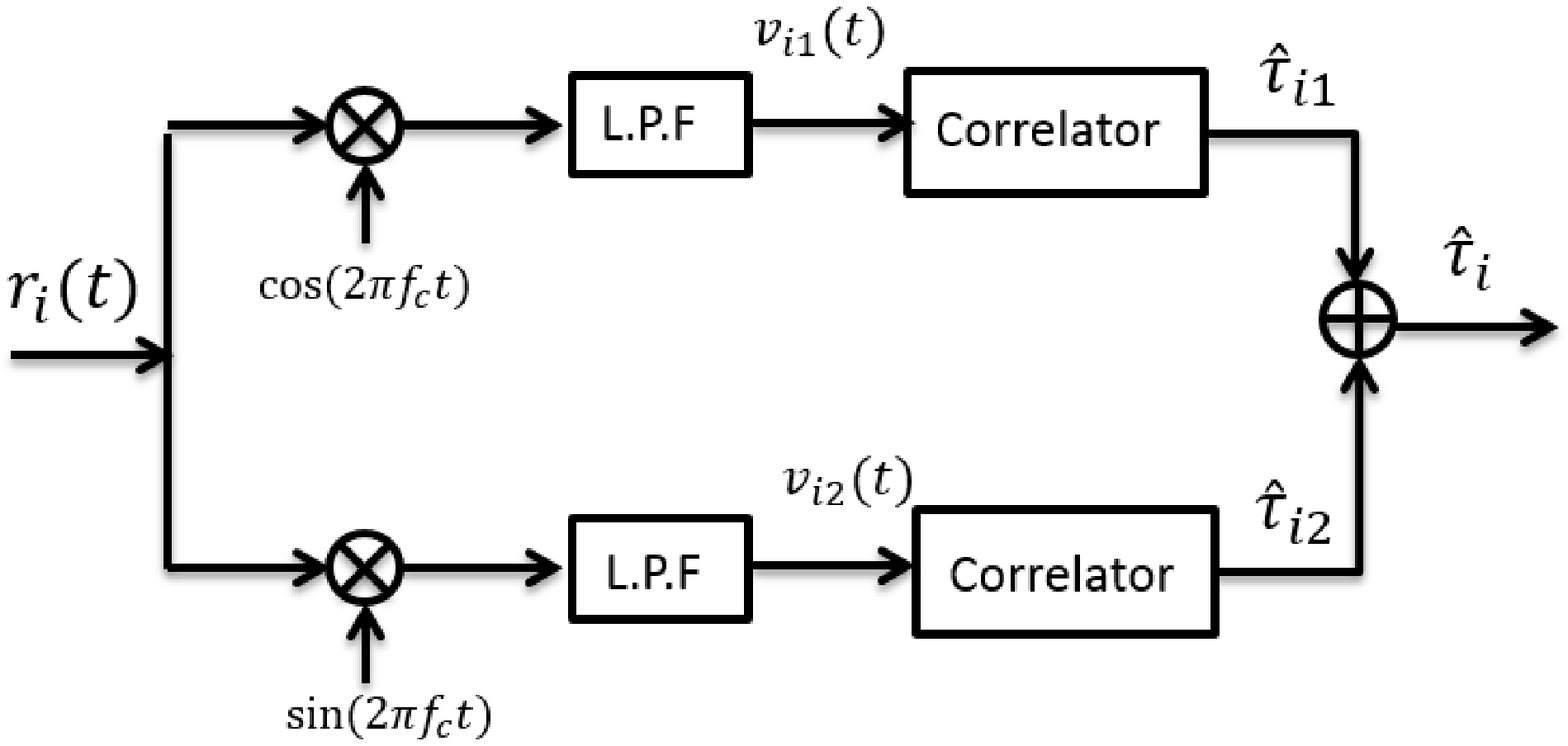}
	}
	\caption{Non-coherent TOA estimation schemes.}
	\label{Fig:TOAestimation}
\end{figure*}

Consider a carrier modulated signal with carrier frequency $f_{c}$ transmitted on a fading channel for TOA estimation. When the received phase is known at each anchor, a coherent estimation strategy, as shown in Figure \ref{fig:coherent} is applied to estimate the TOA. The received signal is given by
\begin{equation}
\label{Eq:receivedsignal}
r_{i}(t)=|h_{i}|\sum_{n}s[n]g(t-nT-\tau_{i}) \cos(2\pi f_{c}t+\theta_{i}),
\end{equation}
is multiplied by $\cos(2\pi f_{c}t+\theta_{i})$ and then low pass filtered. The output of the low pass filter
\begin{equation}
\label{Eq:outputsignal}
v_{i}(t)= \frac{|h_{i}|}{2}\sum_{n}s[n]g(t-nT-\tau_{i})
\end{equation}
is correlated with a regenerated template signal  
\begin{equation}
\label{Eq:corr}
s_{i}(t) = \sum_{n}s[n]g(t-nT-\tau^{*})
\end{equation}
with delay $\tau^{*}$. The TOA is estimated by finding the maximum value of the output of the correlator.

When the phases are unknown, a non-coherent estimation strategy is needed. We will consider non-coherent architectures that correlate with a base-band signal. Figure \ref{Fig:TOAestimation} shows two such non-coherent estimation schemes. Figure \ref{fig:noncoherent} correlates the received signal with a regenerated modulated signal and its 90 degree shifted regenerated signal. In this scheme, the input of the correlator is the sum of the output of two low pass filters, which is
\begin{align}
\label{Eq:outputsignal2}
v_{i}(t)=\frac{1}{2}(|h_{i}|\cos(\theta_{i})- & |h_{i}|\sin(\theta_{i})) \nonumber \\ 
& \sum_{n}s[n]g(t-nT-\tau_{i}).
\end{align}
Similar to the coherent estimation scheme, $v_{i}(t)$ in (\ref{Eq:outputsignal2}) is correlated with the signal given in (\ref{Eq:corr}) to estimate TOA. An alternate non-coherent estimation scheme is shown in Figure \ref{fig:noncoherent2}. In this scheme, in-phase and quadrature components estimate the TOA independently. First, the received signal $x_{i}(t)$ in (\ref{Eq:receivedsignal}) is multiplied separately by $\cos \left(2\pi f_{c}t\right)$ and $\sin \left(2\pi f_{c}t\right)$, and then passed to two low pass filters. The output of the two low pass filters are given by 
\begin{equation}
\label{Eq:vi1}
v_{i1}(t) = \frac{|h_{i}|}{2}\cos(\theta_{i})\sum_{n}s[n]g(t-nT-\tau_{i}),
\end{equation}
and
\begin{equation}
\label{Eq:vi2}
v_{i2}(t) = \frac{|h_{i}|}{2}\sin(\theta_{i})\sum_{n}s[n]g(t-nT-\tau_{i}).
\end{equation}
Then $v_{i1}(t)$, which contains the in-phase component, and $v_{i2}(t)$, which contains the quadrature component, estimate TOA separately by correlating the signal with the regenerated signal that is given in (\ref{Eq:corr}), and two TOA estimates on each branch are given as $\HTaua$ and $\HTaub$ respectively. The final TOA estimate $\HTau$ can be computed by combing $\HTaua$ and $\HTaub$ using different combing methods. The CRLB comparisons between these non-coherent estimation schemes will be provided in Section \ref{sec:nocsi_estimation}. 

To summarize our notation, we denote $\HTau$ as  the estimated time delay between a node and the $i^{th}$ anchor, $d_{i}$ is the true distance between the node and the $i^{th}$ anchor, and $\Sigma$ is the variance of TOA measurements in the absence of fading. In addition, $h_{i}$ is the fading coefficient at the $i^{th}$ channel, and $|h_{i}|$ is the magnitude of $h_{i}$. Meanwhile, $\z$ is the node location, and $\z_{i}$ is the $i^{th}$ anchor location.  In general, a bold letter, for example $\mathbf{x}$, stands for  a vector or matrix, and $||\mathbf{x}||_{2}$ denotes the Euclidean norm of a vector $\mathbf{x}$. Also, $Q(\cdot)$ is the error function.

\subsection{Fading coefficients are known at each anchor}
\label{sec:estimation_known}
Assume $h_{i}$ is known at each anchor. Since both amplitude and phase are known, a coherent estimation strategy is used for location estimation as in Figure \ref{fig:coherent}. Conditioned on the fading coefficients, the TOA measurement $\HTau$ in (\ref{Eq:Gassian}) is Gaussian distributed, and is given by
\begin{equation}
\HTau \sim \mathcal{N}\left(\frac{\d}{c}, \frac{\Sigma}{|h_{i}|^{2}}\right),
\label{Eq:conditionfading}
\end{equation}
where $\text{E}\left[|h_{i}|^{2}\right]=1$. In this case, the CRLB can be expressed as a function of the fading coefficients, with analysis very similar to the case with only additive white Gaussian noise (AWGN)\cite{patwari}:
\begin{equation}
\text{CRLB}_{\text{1-D}}=\frac{c^{2}\Sigma}{\sum^{M}_{i=1}|h_{i}|^{2}}.
\label{Eq:CRB_conditionedfading}
\end{equation} 
Recall that the CRLB in 1-D in the absence of fading \cite{Neal03} is a special case of (\ref{Eq:CRB_conditionedfading}) with $\absa=1$, and is given as
\begin{equation}
\label{Eq:CRLB_AWGN_1D}
\text{CRLB}_{\text{1-D}}^{\text{AWGN}} = \frac{c^{2}\Sigma}{M}.
\end{equation}
Similarly, we can also calculate the CRLB where the fading coefficients are known at each anchor in 2-D. Note that in 2-D, the CRLB depends on the geometry of the network, and it is more complicated than the 1-D case. However, a similar conclusion as the 1-D case that when $\absa=1$, $\text{CRLB}_{\text{2-D}}=\CRLBAWGN2$, can be reached when compared with the AWGN case in \cite{Neal03}.

\subsection{Effect of unknown fading amplitude}
\label{sec:mcrlb}
When the amplitude of fading coefficients is unknown at any anchor, we will show that the presence of fading always degrades the CRLB. To show this, we use the modified CRLB (MCRLB) \cite{MCRLB}, which is defined as  
\begin{equation}
\MCRLB = \text{tr}\left(\left(-\text{E}_{\T,\h}\left[\nabla_{\z}^{2}\text{ln}f(\T|\h,\z)\right]\right)^{-1}\right),
\label{Eq:CRLB_MCRLB}
\end{equation}
where $\nabla_{\z}^{2}$ is the Hessian operator, $ \text{tr}\left(\mathbf{A}\right)$ is the trace of the matrix $\mathbf{A}$, $\mathbf{h}=[|h_{1}|,|h_{2}|,\dots,|h_{M}|]$ contains the amplitude of the fading coefficients, $\T=[\hat{\tau}_{1},\hat{\tau}_{2},\dots,\hat{\tau}_{M}]$ contains all TOA measurements, and $\z$ is the location of the node. 
In one dimension, using (\ref{Eq:conditionfading}), (\ref{Eq:CRLB_MCRLB}) can be calculated as 
\begin{equation}
\text{MCRLB}_{\text{1-D}}=\frac{c^{2}\Sigma}{\sum^{M}_{i=1}\text{E}\left[|h_{i}|^{2}\right]}.
\label{Eq:MCRLB_FINAL}
\end{equation}
Since $\text{E}\left[|h_{i}|^{2}\right]=1$, the MCRLB in  (\ref{Eq:MCRLB_FINAL}) can be expressed as $\text{MCRLB}_{\text{1-D}}=c^{2}\Sigma/M=\text{CRLB}_\text{1-D}^{\text{AWGN}}$, and the MCRLB for the localization error equals to the AWGN case in (\ref{Eq:CRLB_AWGN_1D}), which is also seen in (\ref{Eq:CRB_conditionedfading}) with $|h_{i}|=1$. Since the MCRLB is known to be a lower bound on the CRLB in the presence of fading \cite{MCRLB}, we can conclude that the presence of fading will always degrade the performance for $\it{any}$ fading amplitude distribution. For the MCRLB in 2-D, the derivation is very similar as 1-D, and it turns out the MCRLB in 2-D is the same as the CRLB of the 2-D AWGN case as well. The details are omitted for brevity.

\subsection{Unknown fading amplitude: Nakagami fading}
\label{sec:crlb}
Having seen that fading degrades the performance, we quantify this degradation in the Nakagami envelope case. We assume that fading does not change during the TOA measurements, the phases of the fading coefficients are known at each anchor, and the amplitudes $\absa$ are Nakagami distributed, corresponding to a Gamma distributed $\absaa$. Since the phase is known, the coherent estimation strategy which is used in Section \ref{sec:estimation_known} can be applied. The TOA measurements $\HTau$ are assumed to be i.i.d., and conditioned on the fading coefficients satisfy (\ref{Eq:conditionfading}), where the fading power is Gamma distributed and given by \cite{wireless}:
\begin{equation}
\dh=m^{m}x^{m-1}\Gamma(m)^{-1}\text{exp}(-mx),
\label{Eq:exponential}
\end{equation}
where $\frac{1}{2} \leq m < \infty$ is the Nakagami fading parameter, and as before, $\text{E}\left[|h_{i}|^{2}\right]=1$. When $m=\frac{1}{2}$, the envelope $\absa$ is one-sided Gaussian distributed; when $m=1$, $|h_{i}|$ follows the Rayleigh distribution; and as $ m \rightarrow \infty$, the channel exhibits no fading corresponding to an AWGN channel.

The unconditional distribution of $\HTau$ can be calculated by using the total probability theorem:
\begin{equation}
\df=\int^{\infty}_{0}f\left(\HTau\Big||h_{i}|^{2},\z\right)\dh dx.
\label{Eq:formula}
\end{equation}
By substituting (\ref{Eq:conditionfading}) and (\ref{Eq:exponential}) into (\ref{Eq:formula}), and using \cite[p.310]{tables} we obtain 
\begin{equation}
\df=\frac{m^{m}(m-\frac{1}{2})!}{\sqrt{2 \pi \Sigma}\Gamma(m)\left(\frac{1}{2\Sigma}(\hat{\tau}_{i}- \frac{\d}{c})^{2}+m\right)^{(m+\frac{1}{2})}}.
\label{Eq:Fadingknown}
\end{equation}

For convenience, let $\lz=\text{ln} \df$ be the log likelihood function of each TOA measurement. Due to the independence of the TOA measurements, we define $l(\T|\z)=\sum^{M}_{i=1}\text{ln}\df$. The CRLB can be expressed as \cite{vantrees}
\begin{equation}
\text{CRLB}(\z)=\text{tr}\left(\f^{-1}\right),
\end{equation}
where $\f=-\text{E}_{\T}\left[\nabla_{\z}^{2}l(\T|\z)\right]$ is the Fisher information matrix (FIM). We can calculate the $(j,k)$ element of $\f$, denoted by $[\f]_{jk}$

\begin{equation}
\label{Eq:FI}
[\f]_{jk}=\left \{
\begin{array}{ll}      
\sum^{M}_{i=1}\text{E}_{\HTau} \left[\left( \frac{\partial l(\HTau|\z)}{\partial \z_{j}} \right)^{2} \right] & j=k \\
-\text{E}_{\HTau} \left[ \frac{\partial^{2} l(\HTau|\z)}{\partial \z_{j}\partial \z_{k}} \right] & j\neq k
\end{array}
\right..
\end{equation}

In 1-D, using (\ref{Eq:FI}) and  (\ref{Eq:Fadingknown}), $ \text{E}_{\HTau} \left[ \left( \frac{\partial \lz}{\partial z} \right)^{2} \right]$ can be calculated as
\begin{equation}
\text{E}_{\HTau} \left[ \left( \frac{\partial \lz}{\partial z} \right)^{2} \right] = \frac{m^{m}(m-\frac{1}{2})!(m+\frac{1}{2})^{2}}{\Gamma(m)\sqrt{2\pi}c^2\sigma^{5}}\X 
\label{FI_1d}
\end{equation}
where 
\begin{equation}
\X =\mathlarger{\int^{\infty}_{0}}\frac{(\HTau-\frac{d_{i}}{c})^{2}}{\left( \frac{1}{2\Sigma}(\HTau-\frac{d_{i}}{c})^{2}+m \right)^{\frac{5}{2}+m}}d\HTau.
\label{Eq:X_ori}
\end{equation}
Unlike the AWGN case, the Fisher information depends on $d_{i}$ through $\X$ in (\ref{Eq:X_ori}). However, using \cite[p.292]{tables}, it is possible to express it as 
\begin{align}
\X \leq & \frac{ \sqrt{2}\sigma^{3}\Gamma(\frac{3}{2})\Gamma(m+1)}{m^{1+m}\Gamma(m+\frac{5}{2})} \nonumber \\ 
& +
\frac{\left(\frac{d_{i}}{c}\right)^{2}}{\left(\frac{1}{2\Sigma}(\frac{d_{i}}{c})^{2}+m\right)^{\frac{5}{2}+m}}.
\label{Eq:X_simp}
\end{align}
Since the second term in (\ref{Eq:X_simp}) is small, it is clear that $\X$ can be approximated by the first term, and therefore approximately independent of $d_{i}$. The exact CRLB in the presence of Nakagami fading in 1-D can be expressed as
\begin{equation}
\CRLBr_{\text{1-D}}(\mathbf{z})=\frac{\Gamma(m)\sqrt{2\pi}c^2\sigma^{5}}{m^{m}(m-\frac{1}{2})!(m+\frac{1}{2})^{2}\sum^{M}_{i=1}\X},
\label{CRLB_1d}
\end{equation} 
with an approximation as 
\begin{equation}
\label{Eq:approx1dcrlb}
\CRLBr_{\text{1-D}}(\mathbf{z}) \approx \frac{2c^{2}\Sigma\Gamma\left(m+\frac{5}{2}\right)}{\left(m-\frac{1}{2}\right)!\left(m+\frac{1}{2}\right)^{2}}\frac{1}{M}.
\end{equation}
The approximation of the loss due to fading can be expressed as
\begin{equation}
\frac{\text{CRLB}_{\text{1-D}}(\z)}{\text{CRLB}^{\text{AWGN}}_{\text{1-D}}} \approx k=\frac{\sqrt{\pi} \Gamma(m+\frac{5}{2})}{\Gamma(\frac{3}{2})(m+\frac{1}{2})^{2}\left(m-\frac{1}{2}\right)!},
\label{Eq:k}
\end{equation}
where we recall from (\ref{Eq:CRLB_AWGN_1D}) that $\text{CRLB}^{\text{AWGN}}_{\text{1-D}}=c^{2}\Sigma/M$. As $m \rightarrow \infty$, the second term in (\ref{Eq:X_simp}) goes to $0$ and $k$ in (\ref{Eq:k}) goes to $1$ so that the CRLB in the presence of fading converges to the AWGN case.

When $m=1$, the fading follows the Rayleigh distribution, and the exact CRLB in (\ref{CRLB_1d}) is simplified as
\begin{equation}
\CRLBr_{\text{1-D}}(\z)=\frac{8\sqrt{2}c^{2}\sigma^{5}}{9\sum^{M}_{i=1} \X}.
\label{Eq:rayleigh1d}
\end{equation}
To simplify even further, we use the first term of (\ref{Eq:X_simp}) because $\frac{d_{i}}{c} \approx 0$ and set $m=1$ to obtain 
\begin{equation}
\CRLBr_{\text{1-D}}=\frac{\Sigma c^{2}}{M}\frac{10}{3}.
\label{Eq:comparison_1D}
\end{equation} 
This shows that the loss in SNR due to Rayleigh fading is a factor of $k=\frac{10}{3}$ which is about $5$dB, compared to the AWGN case.

In 2-D, the distance between the node and the $i^{th}$ anchor is $d_{i}=\sqrt{(x_{i}-x)^{2}+(y_{i}-y)^{2}}$. Letting $Y(m)=m^{m}(m-\frac{1}{2})!(m+\frac{1}{2})^{2}\left[\Gamma(m)\sqrt{2\pi}\sigma^{3}\right]^{-1}$. The FIM is
\begin{equation}
\f = \frac{Y(m)}{c^{2}\Sigma}\sum^{M}_{i=1}\X
\begin{bmatrix}
\frac{(x_{i}-x)^{2}}{d_{i}^2}& \frac{(y_{i}-y)(x_{i}-x)}{d_{i}^2} \\
\frac{(y_{i}-y)(x_{i}-x)}{d_{i}^2} & \frac{(y_{i}-y)^{2}}{d_{i}^2} 
\end{bmatrix}.
\label{FIM_2d}
\end{equation}
The CRLB on the variance of the localization error in 2-D is
\begin{equation}
\CRLBr_{\text{2-D}}(\mathbf{z})=\text{tr}\left(\f^{-1}\right).
\label{Eq:CRLBr}
\end{equation}

The FIM in the absence of fading for the 2-D case is given in \cite{Neal03}, and can be written the same as (\ref{FIM_2d}) except without the $Y(m)$ and $\X$ terms. Comparing (\ref{Eq:CRLBr}) with the CRLB in the absence of fading in \cite{Neal03}, both CRLBs in 2-D depend on the true location of the node. When $\frac{d_{i}}{c} \approx 0$, similar to the 1-D case, $\X$ in (\ref{Eq:X_simp}) can be simplified. After simplifications and substituting into (\ref{Eq:CRLBr}), we see that the CRLB in the presence of fading is also a factor of $k$ higher than the AWGN counterpart, i.e. when $m=1$, $k=\frac{10}{3}$ in both 1-D and 2-D. Further, as $m \rightarrow \infty$, the CRLB in 2-D converges to the AWGN case.

\subsubsection*{Extension to multiple nodes case} 
When $N$ nodes exist in a WSN, $\f$ becomes a $N \times N$ matrix, and the diagonal elements in (\ref{Eq:FI}) is summed from $i=1$ to $i=M+N-1$. Using the approximation of $\X$ in (\ref{Eq:X_simp}), after simplifications, in 1-D, the CRLB for the $i^{th}$ node is the $(i,i)$ element of $\f^{-1}$, which is given by
\begin{equation}
\CRLBr_{\text{1-D}}(\mathbf{z}_{i}) \approx \frac{2c^{2}\Sigma\Gamma\left(m+\frac{5}{2}\right)}{\left(m-\frac{1}{2}\right)!\left(m+\frac{1}{2}\right)^{2}}\frac{M+1}{M\left(N+M\right)}.
\label{Eq:CRLB_ext}
\end{equation}
We can prove that in cooperative WSNs, the ratio of location estimation in the presence of fading and in the absence of fading keeps the same.

\subsubsection*{Effect of anchor location}
\label{sec:diffVar}
When anchors are not equidistant from the target node, TOA estimates may have different variances. The CRLB in this case is
\begin{equation}
\label{Eq:CRLB_different_sigma_ori}
\CRLBr_{\text{1-D}}(\mathbf{z}_{i})=\sum^{M}_{i=1}\X\frac{\Gamma(m)\sqrt{2\pi}c^{2}\Sigmai^{5}}{m^{m}(m-\frac{1}{2})!\left(m+\frac{1}{2}\right)^{2}},
\end{equation}
where $\Sigmai$ is the variance of the TOA measurement between the $i^{th}$ anchor and the node. Using a similar approach as the equal variance case, one can prove that when the variance is different among anchors, the ratio of the location estimation in the presence of fading and in the absence of fading is the same as our analysis before.

\subsubsection*{ ML estimator in the presence of Nakagami fading}
\label{sec:ml}
The ML estimator for location estimation in the presence of fading is denoted as
\begin{equation}
\hz=\argmax_{\z}\prod_{i=1}^{M}\df.
\label{Eq:ML}
\end{equation}
Substituting (\ref{Eq:Fadingknown}) into (\ref{Eq:ML}), we have 
\begin{equation}
\hz=\argmin_{\z}\sum^{M}_{i=1}\ln\left(\frac{1}{2\Sigma m}\left(\HTau-\frac{d_{i}}{c}\right)^{2}+1\right),
\label{Eq:ML_final}
\end{equation}
where $d_{i}=||\p_{i}-\z||_{2}$.

In the absence of fading, the ML estimator which is derived in \cite{Neal03} is
\begin{equation}
\hz=\arg\min_{\z}\sum^{M}_{i=1}\left(\HTau-\frac{d_{i}}{c}\right)^{2},
\label{Eq:ML_awgn}
\end{equation}
which is different from (\ref{Eq:ML_final}). Since $\ln(1+x) \approx x$ for small $x$, it is straightforward to see that if $m$ is large, (\ref{Eq:ML_final}) and (\ref{Eq:ML_awgn}) are approximately the same.

\subsection{No CSI available at anchors}
\label{sec:nocsi_estimation}
In the previous sections, we assumed that the phases of the fading coefficients are known at each anchor. When there is no CSI (phase or amplitude) available at any anchor, a non-coherent estimator is applied. Since the optimal non-coherent estimator is hard to implement, one of the suboptimal non-coherent estimators shown in Figure \ref{Fig:TOAestimation} can be applied. When the non-coherent estimator in Figure \ref{fig:noncoherent} is applied, using (\ref{Eq:outputsignal2}) and \cite[p.233]{Helstrom}, conditioned on amplitudes and phases of the fading coefficients, the pdf of the TOA measurements is Gaussian with mean and variance given by
\begin{equation}
\label{Eq:pdf_toa_noCSI}
\HTau \sim \mathcal{N} \left( \frac{\d}{c}, \frac{\Sigma}{|h_{i}|^{2} \left(1-\sin\left( 2\ti \right)\right)}\right).
\end{equation}  
As in Section \ref{sec:crlb}, we assume that $|h_{i}|^{2}$ is Gamma distributed. In addition, we assume $\ti$ is uniformly distributed over $\left[0, 2\pi\right)$, and is independent of $|h_{i}|^{2}$. We can calculate the unconditional distribution of $\HTau$ by integrating the effect of $|h_{i}|^{2}$ and $\ti$, which is given by
\begin{equation}
\label{Eq:pdf_unconditioned_toa_theta_h}
\begin{split}
\df=\int^{2\pi}_{0}\int^{\infty}_{0}f\left(\HTau\Big|\left(|h_{i}|^{2},\ti\right),\z\right) \times \\
\dh\dt dxd\ti,
\end{split}
\end{equation} 
where $\dt=\frac{1}{2\pi}$, $\theta_{i} \in \left[0, 2\pi\right)$, and $\dh$ is given in (\ref{Eq:exponential}).
After simplifications, 
\begin{equation}
\begin{split}
\label{Eq:pdf_unconditioned_toa_theta_h_simp}
\df = \frac{2\sigma m^{m}\left(m-\frac{1}{2}\right)!m^{\frac{1}{2}-m}}{\Gamma(m)\pi^{\frac{3}{2}}\left(\left(\HTau-\frac{d_{i}}{c}\right)^{2}+m\Sigma\right)} \\
\times {_2}F_{1}\left(1, 1-m, \frac{3}{2}; \frac{\left(\HTau-\frac{d_{i}}{c}\right)^{2}}{\left(\HTau-\frac{d_{i}}{c}\right)^{2}+m\Sigma}\right),
\end{split}
\end{equation}
where $\left(a\right)_{n} =a\left(a+1\right) \dots \left(a+n-1\right), n>1$ is the Pochhammer symbol with $\left(a\right)_{0}=1$, and we use  ${_2}F_{1}(a,b,c;z)=\sum^{\infty}_{n=0}\frac{\left(a\right)_{n}\left(b\right)_{n}}{\left(c\right)_{n}}\frac{z^{n}}{n!}$, the hypergeometric function \cite{tables}.

When the amplitude of the fading coefficients is Rayleigh distributed, which means $m=1$, (\ref{Eq:pdf_unconditioned_toa_theta_h_simp}) can be simplified:
\begin{equation}
\label{Eq:pdf_no_csi_sim}
\df = \frac{\sigma}{\pi\left(\Sigma+\left(\HTau-\frac{d_{i}}{c}\right)^{2}\right)},
\end{equation}
which, interestingly, is the Cauchy distribution with scale factor $\sigma$, and median $\frac{d_{i}}{c}$.

Using (\ref{Eq:FI}), in 1-D, the Fisher information can be expressed as
\begin{equation}
\label{Eq:FI_no_csi_1D}
\fz = \frac{4}{c^{2}\sigma^{5}\pi} \sum_{i=1}^{M}Y\left(d_{i}\right),
\end{equation}
where
\begin{align}
\label{Eq:Y(d)}
Y & \left(d_{i}\right) \nonumber \\ 
& = \mathlarger{\int^{\infty}_{0}}\left(\HTau-\frac{d_{i}}{c}\right)^{2}\left(1+\frac{\left(\HTau-\frac{d_{i}}{c}\right)^{2}}{\Sigma}\right)^{-3}d\HTau.
\end{align}
Similar to $X(d_{i})$ in (\ref{Eq:X_ori}) and (\ref{Eq:X_simp}), it is possible to express $Y(d_{i})$ as 
\begin{equation}
\label{Eq:Y(d)_sim}
Y\left(d_{i}\right) \leq \frac{\sigma^{3}\pi}{16}+ \left(\frac{d_{i}}{c}\right)^{2}
\left(1+\frac{\left(\frac{d_{i}}{c}\right)^{2}}{\Sigma}\right)^{-3}.
\end{equation}
Since the second term in (\ref{Eq:Y(d)_sim}) is small, it is clear that $Y(d_{i})$ can be approximated by using the first term, which is independent of $d_{i}$. After simplification, 
the CRLB in 1-D when no CSI is available is 
\begin{equation}
\label{Eq:nocsi_crlb_1d}
\CRLBr_{\text{1-D}}\cong \frac{4c^{2}\Sigma}{M}.
\end{equation}
Recalling (\ref{Eq:comparison_1D}), we see that when no CSI is available at any anchor, the loss in SNR is a factor of $k=4$, which is about $6$dB. To calculate the CRLB in 2-D we can use (\ref{Eq:FI}) to calculate the elements of the FIM. Similar to the 1-D case, the loss in SNR compared to the AWGN case is also $6$dB.

When the alternate non-coherent estimator in Figure \ref{fig:noncoherent2} is applied, conditioned on the amplitudes and phase of the fading coefficients, the distribution of $\HTaua$ and $\HTaub$ can be obtained using \cite[p.233]{Helstrom}, (\ref{Eq:vi1}) and (\ref{Eq:vi2}) as
\begin{equation}
\HTaua \sim \mathcal{N} \left( \frac{\d}{c}, \frac{\Sigma}{|h_{i}|^{2} \cos^{2}\left(\ti \right)}\right),
\label{Eq:HTaua}
\end{equation} 
and
\begin{equation}
\HTaub \sim \mathcal{N} \left( \frac{\d}{c}, \frac{\Sigma}{|h_{i}|^{2} \sin^{2}\left(\ti \right)}\right).
\label{Eq:HTaub}
\end{equation} 
Since $\ti$ is uniformly distributed, both $\cos^{2}(\ti)$ and $\sin^{2}(\ti)$ have the same distribution. Therefore we will focus on $\HTaua$. Using the formula $\cos^{2}\ti=\frac{1}{2}\left(1-\cos 2 \ti\right)$, and the fact that $\frac{1}{2}\left(1-\cos 2\ti\right)$ has the same distribution as $\frac{1}{2}\left(1-\sin 2\ti\right)$ when $\ti$ is uniformly distributed, comparing (\ref{Eq:HTaua}) with (\ref{Eq:pdf_toa_noCSI}) one can see that the variance of (\ref{Eq:HTaua}) is twice of (\ref{Eq:pdf_toa_noCSI}). Therefore, the CRLB when the quadrature component is extracted is $2$ times higher than the CRLB when the previous non-coherent estimation scheme is applied. Also, since the unconditional distribution of $\HTaua$ is the same as $\HTaub$, the CRLB when the in-phase component is extracted is the same as the CRLB when the quadrature component is extracted. If the average is taken between $\HTaua$ and $\HTaub$, $\HTau=\frac{\HTaua+\HTaub}{2}$ the final CRLB is the same as the previous non-coherent estimation scheme, which is given in (\ref{Eq:nocsi_crlb_1d}).

Even though the noncoherent architectures in Figure \ref{fig:noncoherent} and Figure \ref{fig:noncoherent2} have the same performance, the two schemes have advantages and disadvantages. On the one hand, when there is some prior information on the TOA measurement, the scheme shown in Figure \ref{fig:noncoherent2} is more flexible when combing the two estimates, and therefore can give a better performance. For example, if a range for the TOA measurement is known, between $\HTaua$ and $\HTaub$, the one within the range can be chosen as the final $\HTau$. On the other hand, the scheme shown in Figure \ref{fig:noncoherent} is less complex than the scheme shown in Figure \ref{fig:noncoherent2}, since the latter scheme requires two correlators.

\subsubsection*{ML estimator when no CSI is available at anchors}
When no CSI is available at any anchor, assuming $m=1$ and using (\ref{Eq:ML}) and (\ref{Eq:pdf_no_csi_sim}), the ML estimator for the location estimate is given by
\begin{equation}
\label{Eq:ML_noCSI}
\hz=\argmin_{\z}\sum^{M}_{i=1}\ln \left[\frac{1}{\Sigma}\left(\HTau-\frac{d_{i}}{c}\right)^{2}+1\right].
\end{equation}
Consider a comparison of (\ref{Eq:ML_noCSI}) with (\ref{Eq:ML_final}), which is the ML estimator for the case with known phase and Nakagami envelope. Setting $m=1$ in (\ref{Eq:ML_final}) we see that the only difference between these two ML estimators is a factor of $2$ multiplying $\Sigma$. If we write (\ref{Eq:ML_noCSI}) as a function of $\T=[\hat{\tau}_{1},\hat{\tau}_{2},\dots,\hat{\tau}_{M}]$ and $\Sigma$ as $\hz = g\left(\T, \Sigma\right)$, then $\hz$ in (\ref{Eq:ML_final}) can be expressed as $\hz=g\left(\T, 2\Sigma\right)$. This indicates that the ML estimator with no CSI needs $3$dB higher SNR to use the exact same location estimator as the ML estimator which knows the phases. Note that this does not mean that the performance of (\ref{Eq:ML_final}) and (\ref{Eq:ML_noCSI}) are $3$dB apart, since the distribution of $\T$ in the two cases is different.

Interestingly, comparing the pdf of the TOA measurements with phase information, given in (\ref{Eq:Fadingknown}), with the pdf of TOA measurements with no CSI information, given by (\ref{Eq:pdf_no_csi_sim}), one can see that setting $m=\frac{1}{2}$ in (\ref{Eq:Fadingknown}) is identical to (\ref{Eq:pdf_no_csi_sim}) when $m=1$. Recalling that $m=\frac{1}{2}$ represents the worst Nakagami fading scenario, we conclude that with phase information, the coherent estimation with $m=\frac{1}{2}$ (worse fading) has identical pdf and performance as the non-coherent estimation with $m=1$, i.e. under a better fading scenario.

\subsection{Extension to cooperative location estimation in the presence of fading}
In the previous sections, we only considered a sensor network with $1$ node and $M$ anchors. However, the results can be extended to a cooperative location estimation problem. In this section, we consider a sensor network with $N$ nodes and $M$ anchors, and we assume all nodes communicate with each other. When the fading coefficients are random with Nakagami distributed amplitude, in 1-D, $\text{E}_{\HTau} \left[\left( \frac{\partial l(\HTau|\z)}{\partial \z_{j}} \right)^{2} \right]$ is given in (\ref{FI_1d}), however, $[\f]_{jk}$ is given as
\begin{equation}
\label{Eq:coop-FI-1d}
[\f]_{jk}=\left \{
\begin{array}{ll}      
\sum^{M+N-1}_{i=1}\text{E}_{\HTau} \left[\left( \frac{\partial l(\HTau|\z)}{\partial \z_{j}} \right)^{2} \right] & j=k \\
-\text{E}_{\HTau} \left[ \frac{\partial^{2} l(\HTau|\z)}{\partial \z_{j}\partial \z_{k}} \right] & j\neq k
\end{array}
\right.
\end{equation}
Comparing (\ref{Eq:coop-FI-1d}) with (\ref{Eq:FI}), due to the cooperation between nodes, when $j=k$, each node receives information from other nodes as well. Therefore, when $j=k$, (\ref{Eq:coop-FI-1d}) contains $M+N-1$ terms. Using the first term of $\X$ in (\ref{Eq:X_simp}), and after simplification, we have 
\begin{equation}
\label{Eq:coop-FI-1d-ex}
\text{E}_{\HTau} \left[\left( \frac{\partial l(\HTau|\z)}{\partial \z_{j}} \right)^{2} \right]=\frac{\Gamma\left(m+\frac{1}{2}\right)!\left(m+\frac{1}{2}\right)^{2}}{2\Gamma\left(m+\frac{5}{2}\right)c^{2}\Sigma}.
\end{equation}
The CRLB for the $i^{th}$ node is the $(i,i)$ element of $\mathbf{F}^{-1}_{\mathbf{z}}$, which is given by
\begin{align}
\label{Eq:coop-1d-ii}
\text{CRLB}_{\text{1-D}}(\textbf{z}_{i}) = & \frac{2\Gamma\left(m+\frac{5}{2}\right)c^{2}\Sigma}{m\Gamma\left(m+\frac{1}{2}\right)\left(m+\frac{1}{2}\right)^{2}} \nonumber \\
& \times \frac{M+1}{M\left(N+M\right)}.
\end{align}
When $m=1$, the ratio between cooperative location estimation in (\ref{Eq:coop-1d-ii}) and non-cooperative location estimation in (\ref{Eq:comparison_1D}) is $R= (M+1)/(N+M)$. Since $N \geq 1$ in a cooperative network, (\ref{Eq:coop-1d-ii}) is always equal or smaller than (\ref{Eq:comparison_1D}), which proves that cooperation between nodes gives a lower CRLB. In 2-D, the similar conclusion can be reached. 

When no CSI is available at each anchor, with the assumption that the amplitude of fading coefficients is Rayleigh distributed, the pdf of the TOA measurements is given in (\ref{Eq:pdf_no_csi_sim}). Therefore in 1-D, the CRLB of the $i^{th}$ node in a cooperative WSN with $N$ nodes and $M$ anchors is
\begin{equation}
\label{Eq:coop-1d-crlb}
\text{CRLB}_{\text{1-D}}(\textbf{z}_{i})=\frac{1}{4c^{2}\Sigma}\frac{M+1}{M(N+M)},
\end{equation}
for which the ratio between the cooperative and the non-cooperative is still $r$. Therefore, we can conclude that cooperation between nodes results in a lower CRLB.

\section{Location Detection in the presence of fading}

\label{sec:detection}
In Section \ref{sec:estimation} we have considered the case that the location of the node is unknown. However, in some applications, the node location is known to all anchors, but whether the node is active or not is unknown. In many applications such as detecting fire in buildings, each node is placed inside a room, and the location is known to all anchors. Anchors detect an event based on whether the node is transmitting. 

Similar to the estimation case, we consider a sensor network with $M$ anchors and one node. In the absence of the node, each anchor receives only noise. In the presence of the node, each anchor receives faded signal with noise. If the phases of the fading coefficients are known at each anchor, a coherent detection scheme, which needs only one phase-synchronized matched filter, can be applied. Similar to the estimation formulation, we assume the TOA measurement is made at each anchor. Since the arrival time can be estimated using both continuous time signal and discrete time signal with high enough sampling frequency \cite{Zekavat11}, in this section, we assume discrete time signals are extracted to detect the presence or absence of transmission. In this case, $v_{i}(t)$ in Figure \ref{fig:coherent} is sampled and a total number of $N$ samples are extracted. Next, the $N$ samples are correlated with the sampled transmitted signal and compared with a threshold. If the phase is unknown at any anchor, a non-coherent detection scheme is applied. In this case, two demodulators with $90$ degree phase shift of each other are needed, which is shown in Figure \ref{fig:noncoherent}. Then $v_{i}(t)$ in Figure \ref{fig:noncoherent} is sampled and $N$ samples are extracted. By comparing the output of correlator with a threshold, a final decision is made at each anchor. If an anchor detects the node, it transmits a bit $ ``1"$ to the FC, otherwise it transmits a bit $ ``0"$. The FC needs at least $K$ anchors to declare the node exists at the given location, where $K$ is a design parameter.

To detect the node, a binary hypothesis testing problem at the $i^{th}$ anchor can be formulated as 
\begin{equation}
\label{eq:system_model_detection}
\rin =
\begin{cases} 
\omega_i[n] & \mathrm{under \:} H_{0}   \\
h_{i}s_{i}[n-n_{i}]+\omega_i[n]  & \mathrm{under \:} H_{1} 
\end{cases}.
\end{equation}
As before the fading coefficients, $h_{i}$, are complex Gaussian random variables. Both real and imaginary parts of $h_{i}$ have $0$ mean and variance $\frac{1}{2}$, to satisfy $\text{E}\left[|h_{i}|^{2}\right]=1$; $\omega_i[n]$ is additive Gaussian noise with $0$ mean and variance $\Sigma$; $s_{i}[n]$ is the modulated deterministic transmitted signal and its total energy $\mathcal{E}=\sum^{N-1}_{n=0}s^{2}_{i}[n]$ is normalized; $n_{i}$ is the true time delay between the node and the $i^{th}$ anchor, where $i=1, 2, \dots, M$. The following three cases are considered in this work. (a) The fading coefficients are assumed to be known at each anchor. In this case, conditioned on the fading coefficients, $r_{i}[n]$ is Gaussian distributed under both $H_{0}$ and $H_{1}$; (b) Amplitudes of the fading coefficients are unknown at any anchor but with a known prior distribution. In this case, the Neyman-Pearson detector can be found by integrating the fading effect \cite{Kay}; (c) No CSI is available at any anchor. In this case, a non-coherent detection scheme which extracts both in-phase and quardrature components is used.

\subsection{Fading coefficients known at anchors}
\label{subsec:det_known}
When the fading coefficients are known at each anchor, the phases can be synchronized, and the coherent detection scheme can be used. The hypothesis testing problem at each anchor can be formulated as 
\begin{equation}
\label{eq:system_model_detection_phase_known}
\rin =
\begin{cases} 
\omega_i[n] & \mathrm{under \:} H_{0}   \\
|h_{i}|s_{i}[n-n_{i}]+\omega_i[n]  & \mathrm{under \:} H_{1} 
\end{cases}.
\end{equation}
Since $h_{i}$ is a zero-mean complex Gaussian random variable, $|h_{i}|$ is Rayleigh distributed. Conditioned on $|h_{i}|$ and based on Neyman-Pearson theorem, the $i^{th}$ anchor detects the node if the likelihood ratio $\Txa$ satisfies
\begin{equation}
\label{Eq:det_Tx}
\Txa = \frac{f\left(\r \Big||h_{i}|;H_{1}\right)}{f\left(\r;H_{0}\right)} \lessgtr \gamma_{i},
\end{equation}
for some threshold $\gamma_{i}$, which balances the false alarm and detection probabilities at each anchor. Here, $\r=[\ri[0], \ri[1],\dots, \ri[N-1]]^{T}$,  $f\left(\r;H_{0}\right)$ is the pdf of the received signal under $H_{0}$, and $f\left(\r \Big||h_{i}|;H_{1}\right)$ is the pdf of the received signal conditioned on $|h_{i}|$ under $H_{1}$. Under both $H_{0}$ and $H_{1}$, $\r$ is Gaussian distributed, given by
\begin{equation}
\label{Eq:det_px_H0}
f\left(\r;H_{0}\right) = \frac{1}{\left(2\pi\Sigma\right)^{\frac{N}{2}}}\exp\left(-\frac{1}{2\Sigma}\sum^{N-1}_{n=0}\ri^{2}[n]\right),
\end{equation}
and
\begin{align}
f & \left(\r\Big||h_{i}|;H_{1}\right) =  \frac{1}{\left(2\pi\Sigma\right)^{\frac{N}{2}}} \nonumber \\ 
& \times \exp\left(-\frac{1}{2\Sigma}\sum^{N-1}_{n=0} \left(\rin-|h_{i}|s_{i}\left[n-n_{i}\right]\right)^{2}\right).
\label{Eq:det_px_H1}
\end{align}

Taking the log of (\ref{Eq:det_Tx}), and substituting (\ref{Eq:det_px_H0}) and (\ref{Eq:det_px_H1}) into (\ref{Eq:det_Tx}), and simplifying, we have

\begin{equation}
\label{Eq:det_Txa}
\ln \Txa = \sum^{N-1}_{n=0}\rin s_{i}\left[n-n_{i}\right] \lessgtr \frac{\Sigma\ln \gamma_{i}}{|h_{i}|} +\frac{|h_{i}|}{2}.
\end{equation}
Define $\Txx=\frac{1}{\sigma} \ln \Txa$, so that (\ref{Eq:det_Txa}) can be expressed as
\begin{equation}
\label{Eq:det_new_threshold}
\Txx \lessgtr \gammaprime\left(\absa\right),
\end{equation}
where 
\begin{equation}
\label{Eq:det_gammaprime}
\gammaprime\left(\absa\right)=\sqrt{\frac{\Sigma}{\absaa}}\ln \gamma_{i}+\frac{1}{2}\sqrt{\frac{\absaa}{\Sigma}}
\end{equation}
is the threshold for the statistic $\Txx$ and depends on both $\gamma_{i}$ and $\absa$. Note that conditioned on $\absa$, $\Txx$ is Gaussian:
\begin{equation}
\label{Eq:det_distTx}
\Txx \sim
\begin{cases} 
\mathcal{N}\left(0, 1\right) & \mathrm{under \:} H_{0} \\
\mathcal{N}\left(\sqrt{\frac{\absaa}{\Sigma}}, 1\right) & \mathrm{under \:} H_{1}
\end{cases}.
\end{equation} 
The instantaneous probability of false alarm at the $i^{th}$ anchor $\pfai$ is 
\begin{equation}
\label{Eq:det_pfaicond}
\pfai = Q\left(\gammaprime\left(\absa\right)\right),
\end{equation}
and the probability of detection at the $i^{th}$ anchor $\pdi$ is 

\begin{equation}
\label{Eq:det_pdicond}
\pdi = Q\left(\gammaprime\left(\absa\right)-\sqrt{\frac{\absaa}{\Sigma}}\right),
\end{equation}
both of which are functions of $\absa$. The average false alarm probability $\pfa$ can be calculated by averaging out the fading effect, which is given by 
\begin{equation}
\label{Eq:det_pfaicond_uncond}
\pfa = \text{E}\left[\pfai\right]=\int^{\infty}_{0}Q\left(\gammaprime\left(\absa\right)\right)f_{|h_{i}|^{2}}\left(x\right)dx,
\end{equation}
where the pdf of $|h_{i}|^{2}$ is given in (\ref{Eq:exponential}).

Similarly, the averaged $\pd$ can be calculated by
\begin{equation}
\label{Eq:det_pdicond_uncond_simp}
\pd = \int^{\infty}_{0}Q\left(\gammaprime\left(\absa\right)-\sqrt{\frac{\absaa}{\Sigma}}\right)f_{|h_{i}|^{2}}\left(x\right)dx.
\end{equation}

After a decision is made at each anchor, it transmits a $ ``1"$ or $ ``0"$ to a fusion center. The fusion center needs to receive at least $K$ $ ``1"$s to decide the node is active, where $K$ is a predetermined design parameter. Therefore, the total probability of false alarm and the total probability of detection are given by

\begin{equation}
\Bpfa =\sum_{m=K}^{M} \binom{M}{m} \left(\pfa\right)^{m}\left(1-\pfa\right)^{M-m},
\label{Eq:det_overallpfa}
\end{equation}
and 
\begin{equation}
\Bpd =\sum_{m=K}^{M} \binom{M}{m} \left(\pd\right)^{m}\left(1-\pd\right)^{M-m}.
\label{Eq:det_overallpd}
\end{equation}

Recall that the threshold $\gammaprime\left(\absa\right)$ given in (\ref{Eq:det_new_threshold}) depends on the random channel amplitude $\absa$. The question arises as to whether this choice is optimal in the sense of maximizing $\pd$ when $\pfa \leq \alpha$, where $\alpha$ is a constant. Now we want to prove that the threshold $\gammaprime\left(\absa\right)$ in (\ref{Eq:det_gammaprime}) has the optimal dependence on $\absa$. We do this by casting the threshold optimization problem as a variational problem where the variable $\gammaprime\left(\absa\right)$ is a function of the channel:
\newtheorem{thm1}{thm1}
\begin{thm}
	\label{thm1}
	\rm{Consider the following optimization problem}
	$$
	\begin{array}{lc}
	\rm{maximize} \\
	\pd = \int^{\infty}_{0}Q\left(\gammaprime\left(\absa\right)-\sqrt{\frac{\absaa}{\Sigma}}\right)f_{|h_{i}|^{2}}\left(x\right)dx \\
	\rm{subject \; to} \\
	\pfa = \int^{\infty}_{0}Q\left(\gammaprime\left(\absa\right)\right)f_{|h_{i}|^{2}}\left(x\right)dx \leq \alpha 
	\end{array}\\,
	$$
	\rm{where the variable function is $\gammaprime\left(\absa\right)$. The optimal threshold function is given in (\ref{Eq:det_gammaprime})}.
\end{thm}
Proof: Please see the Appendix.

\subsection{Fading coefficients with known phase but unknown amplitude}
\label{subsec:det_unknown_amplitude}
Now assume that the amplitude of the fading are unknown at every anchor with a known distribution. In this case, $|h_{i}|$ in (\ref{eq:system_model_detection_phase_known}) is unknown but with a known distribution. The Neyman-Pearson detector at the $i^{th}$ anchor can be formulated as
\begin{align}
\Tx &= \frac{f(\r;H_{1})}{f(\r;H_{0})} \nonumber \\ 
&=\frac{\int^{\infty}_{0}f\left(\r\Big|\absa;H_{1}\right)f_{|h_{i}|}\left(x\right)dx}{f\left(\r;H_{0}\right)} \lessgtr \gamma_{i}. 
\label{Eq:det_det_unconditionfading}
\end{align} 
Defining $\Txx =\frac{1}{\Sigma}\sum^{N-1}_{n=0} \rin s_{i}[n-n_{i}]$ and assume $\absa$ is Rayleigh distributed,  we can express (\ref{Eq:det_det_unconditionfading}) as
\begin{align}
L_{i} & = T_{i} \exp\left(\frac{T_{i}^{2}}{2\left(1+\brho\right)}\right) \nonumber \\ 
& \times \left(1-Q\left(\frac{T_{i}}{\sqrt{1+\brho}}\right)\right) \lessgtr \gamma_{i}^{\prime},
\label{Eq:det_uncond_detector}
\end{align}
where 
\begin{equation}
\label{Eq:gammaprime}
\gamma_{i}^{\prime}=\frac{\left(\gamma_{i}-\frac{1}{\left(1+\brho\right)}\right)\left(1+\brho\right)^{\frac{3}{2}}}{\sqrt{2\pi}},
\end{equation}
and we drop the dependence of $L_{i}$ and $T_{i}$ on $\x$ to emphasize their functional relationship.

We found that for all SNR values, $L_{i}$ is a monotonically increasing function of $T_{i}$. Therefore, we can rewrite (\ref{Eq:det_uncond_detector}) as
\begin{equation}
\label{Eq:det_uncond_detector_highsnr}
\Txx = \sum^{N-1}_{n=0}\rin s_{i}[n-n_{i}] \lessgtr \gamma^{\prime\prime}_{i},
\end{equation}
here $\gamma^{\prime\prime}_{i}$ is a constant, which is not a function of the measured data, and can be calculated numerically as we now explain.
Since the distribution of $\Txx \sim \mathcal{N}\left(0, 1\right)$, $\pfa$ can be calculated as
\begin{equation}
\label{Eq:det_uncond_pfa_nophase}
\pfa = Q\left(\gamma^{\prime\prime}_{i}\right),
\end{equation}
where $\gamma^{\prime\prime}_{i}$ can be found by taking the inverse of (\ref{Eq:det_uncond_pfa_nophase}). 

Since under $H_{1}$ shown in (\ref{Eq:det_distTx}), conditioned on $\absa$, the distribution of $\Txx \sim \mathcal{N} \left(\sqrt{\frac{\absaa}{\Sigma}}, 1\right)$, the detection probability $\pdi$ can be calculated as  
\begin{equation}
\label{Eq:det_pdicond_nophase}
\pdi = Q\left(Q^{-1}\left(\pfa\right)-\sqrt{\frac{\absaa}{\Sigma}}\right).
\end{equation}
The averaged $\pd$ can be calculated using
\begin{equation}
\label{Eq:det_pdicond_uncond_constant}
\pd = \int^{\infty}_{0}Q\left(Q^{-1}\left(\pfa\right)-\sqrt{\frac{\absa}{\Sigma}}\right)f_{|h_{i}|^{2}}\left(x\right)dx,
\end{equation}
which is different from (\ref{Eq:det_pdicond_uncond_simp}), since $\gamma^{\prime\prime}_{i}$ does not depend on $\absa$ anymore. Finally, the total probability of false alarm and detection can be calculated using (\ref{Eq:det_overallpfa}).

Recalling that (\ref{Eq:det_Txa}), is the detector for the case when $h_{i}$ is known to anchors. Comparing it with (\ref{Eq:det_uncond_detector_highsnr}), which is the detector for the case when $|h_{i}|$ is a random variable with Rayleigh distribution, one can see that the detector for both cases rely on the same statistic $\Txx$. However, in (\ref{Eq:det_Txa}) the threshold is a function of the fading coefficients, whereas in (\ref{Eq:det_uncond_detector_highsnr}) the threshold is a constant that only depends on the prescribed $\pfa$. In Section \ref{sec:simulations} we will show (\ref{Eq:det_Txa}) outperforms (\ref{Eq:det_uncond_detector_highsnr}) as expected. The closed form expression for $\pd$ in (\ref{Eq:det_pdicond_uncond_constant}) is not tractable. However, when no CSI is available at each anchor, we will have a closed form expression for $\pd$ as seen next.

\subsection{No CSI is available at any anchor}
\label{sec:No CSI}
When the knowledge of CSI is not available at any anchor, a non-coherent detection scheme is applied. The received bandpass signal is sampled and both in-phase and quadrature components of the signal are extracted \cite{bandpass}. The problem statement can be formulated as
\begin{align}
& \rin= \nonumber \\ 
& \begin{cases} 
\omega_{i}[n] & \mathrm{under \:} H_{0}   \\
h_{\text{R}i}s_{1i}[n-n_{i}] & \\
+h_{\text{I}i}s_{2i}[n-n_{i}]+\omega_{i}[n]  & \mathrm{under \:} H_{1} 
\end{cases}
\label{Eq:det_nocsi_prosta}
\end{align}
where $s_{1i}[n]$ and $s_{2i}[n]$ are the sampled signal $s_{i}[n]$ multiplied by $\cos(2 \pi f_{c}n)$ and $\sin(2 \pi f_{c} n)$ respectively, $f_{c}$ is the carrier frequency, and $\omega_{i}[n]$ can be decomposed into in-phase and quadrature components, $\omega_{\text{R}i}[n]$ and $\omega_{\text{I}i}[n]$ \cite[\S5.4]{Kay}. Similarly, $h_{\text{R}i}$ and $h_{\text{I}i}$ are the real and imaginary parts of the fading coefficients respectively. We can rewrite the hypothesis testing problem in vector format as 
\begin{equation}
\label{Eq:det_nocsi_vec}
\r_{i} =
\begin{cases} 
\boldsymbol{\omega}_{i} & \mathrm{under \:} H_{0}   \\
\mathbf{S}_{i}\mathbf{h}_{i}+\boldsymbol{\omega}_{i}  & \mathrm{under \:} H_{1} 
\end{cases},
\end{equation}
where $\mathbf{x}_{i}=\left[x_{i}[0], x_{i}[1], \dots, x_{i}[N-1]\right]^{T}$, and $\mathbf{S}_{i}$, $\mathbf{h}_{i}$, and $\boldsymbol{\omega}_{i}$ are defined as 
\begin{equation}
\label{Eq:det_nocsi_hi}
\mathbf{S}_{i}=
\begin{bmatrix}
s_{1i}[-n_{i}] & s_{2i}[-n_{i}]\\
s_{1i}[1-n_{i}] & s_{2i}[1-n_{i}] \\
\vdots  & \vdots\\
s_{1i}[N-1-n_{i}] & s_{2i}[N-1-n_{i}] \\ 
\end{bmatrix},
\end{equation}

\begin{equation}
\label{Eq:det_nocsi_thetai}
\mathbf{h}_{i} = \left[h_{\text{R}i}  \quad h_{\text{I}i}\right]^{T}
\end{equation}
and
\begin{equation}
\label{Eq:det_nocsi_omegai}
\boldsymbol{\omega}_{i}=
\begin{bmatrix}
\omega_{\text{R}i}[0] +  \omega_{\text{I}i}[0] \\
\omega_{\text{R}i}[1] +  \omega_{\text{I}i}[1] \\
\vdots \\
\omega_{\text{R}i}[N-1] + \omega_{\text{I}i}[N-1]\\ 
\end{bmatrix}.
\end{equation}

The detector at the $i^{th}$ anchor can be computed by calculating the log likelihood ratio which is a quadratic function of $\r_{i}$ 
\begin{equation}
\label{Eq:det_nocsi_detector}
\Tx = \r_{i}^{T}\mathbf{S}_{i}\mathbf{S}_{i}^{T}\left(\mathbf{S}_{i}\mathbf{S}_{i}^{T}+\Sigma\mathbf{I}\right)^{-1}\r_{i} \lessgtr \gamma_{i},
\end{equation}
After simplifications, $\Tx$ can be expressed as
\begin{equation}
\label{Eq:det_nocsi_detctor_final}
\Tx = \frac{1}{N}\r_{i}^{T}\mathbf{S}_{i}\mathbf{S}_{i}^{T}\r_{i} \lessgtr \gamma_{i}.
\end{equation}

The probability of false alarm and the probability of detection at the $i^{th}$ anchor can be calculated as \cite[\S5.4]{Kay}
\begin{align}
\label{Eq:det_nocsi_pfa}
\pfa &= \text{Pr}\{\Tx>\gamma_{i};H_{0}\} \notag \\
&= \exp\left(-\frac{\gamma_{i}}{12\Sigma}\right).
\end{align}
In addition, the averaged probability of detection $\pd$ can be expressed as a function of the $\pfa$ in closed form:
\begin{align}
\label{Eq:det_nocsi_pd}
\pd &= \text{Pr}\{\Tx>\gamma_{i};H_{1}\} \notag \\
&= \left(\pfa\right)^{\frac{1}{1+\frac{N}{4\Sigma}}}.
\end{align}
The overall $\Bpfa$ and $\Bpd$ can be calculated by substituting (\ref{Eq:det_nocsi_pfa}) and (\ref{Eq:det_nocsi_pd}) into (\ref{Eq:det_overallpfa}) and (\ref{Eq:det_overallpd}).

Comparing different cases, we can see that when no CSI is available at any anchor, it is possible to express the detection probability as a function of the false alarm probability in closed-form as shown in (\ref{Eq:det_nocsi_pd}) at each anchor. In other scenarios, closed-form expressions are not tractable. Therefore, $\pd$ is computed numerically. In all cases, $\Bpfa$ and $\Bpd$ can be computed by substituting $\pfa$ and $\pd$ into (\ref{Eq:det_overallpfa}) and (\ref{Eq:det_overallpd}). 

\subsection{The choice of the design parameter $K$}
\label{sec:K}
As mentioned before, $K$ is a predetermined design parameter which is used to fuse the binary decisions from each anchor in the following manner. If $K$ or more anchors detect a node, a final $detect$ decision is made. When $K$ is large, which means the fusion center requires most of the anchors to claim the node exists, both $\Bpfa$ and $\Bpd$ decrease. On the other hand, When $K$ is small, both $\Bpfa$ and $\Bpd$ increase. However, for a given total false alarm threshold $\Bpfa \leq \alpha$, it appears that there is an optimal value of $K$ which maximizes $\Bpd$. In Section \ref{sec:simulations} we will see the choice of $K$ under different values of $\Bpfa$.

\section{Simulation Results}
\label{sec:simulations}

\subsection{Numerical Results for location estimation}
\label{sec:comparison}
\begin{figure}[t]
	\centering
	\includegraphics[
	height=2.4in,
	width=3in
	]
	{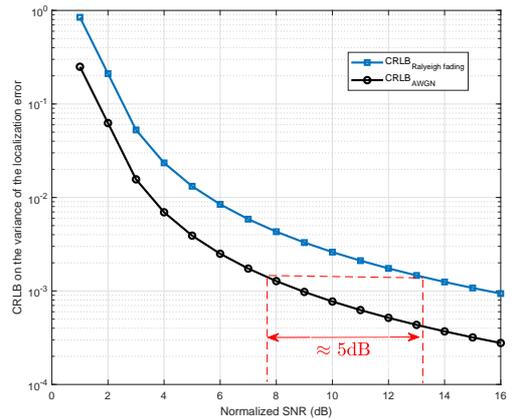}
	\caption{CRLB comparison when SNR is large.}
	\label{fig:infinity}
\end{figure}

We first consider a location estimation problem in 1-D. There are $M=4$ anchors and $1$ node. Figure \ref{fig:infinity} shows the CRLB comparison in 1-D between the AWGN case and the presence of Rayleigh fading. In the high SNR regime, to maintain the same variance of localization error, CRLB in the presence of Rayleigh fading requires about $5$dB more power than the AWGN case.  

\begin{figure}[t]
	\centering
	\includegraphics[
	height=2.4in,
	width=3in
	]
	{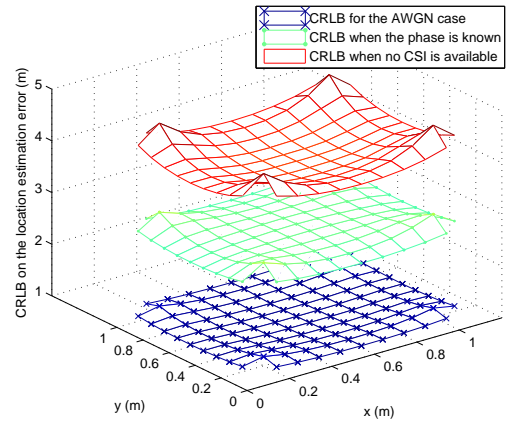}
	\caption{CRLB comparison in a $1\text{m} \times 1\text{m}$ square with $\sigma=\frac{1}{c}$.}
	\label{fig:1m1c}
\end{figure}

Consider now a sensor network with $M=4$ anchors at the corners of a $1\text{m} \times 1\text{m}$ square, and one node within the square, and Rayleigh fading. In Figure \ref{fig:1m1c}, we compare the CRLB for the AWGN case in \cite{Neal03}, the CRLB in (\ref{Eq:CRLBr}), in which the phase information is known at each anchor, and the CRLB in Section \ref{sec:nocsi_estimation} when no CSI is available. We observe that the CRLB for the AWGN case is the lowest, followed by the CRLB when the phase of the fading coefficients is known. When there is no CSI available at any anchor, the CRLB is the highest. In addition, one can see that ratio between $\text{CRLB}_{\text{2-D}}\left(\z\right)$ in (\ref{Eq:CRLBr}) is about  $k=\frac{10}{3}$ higher than the CRLB in the absence of fading, and the CRLB in the presence of fading without knowing the phase at any anchor is a factor of $k=4$ higher than the CRLB in the absence of fading, which corroborates (\ref{Eq:nocsi_crlb_1d}).

\begin{figure}[t]
	\centering
	\includegraphics[
	height=2.4in,
	width=3in
	]
	{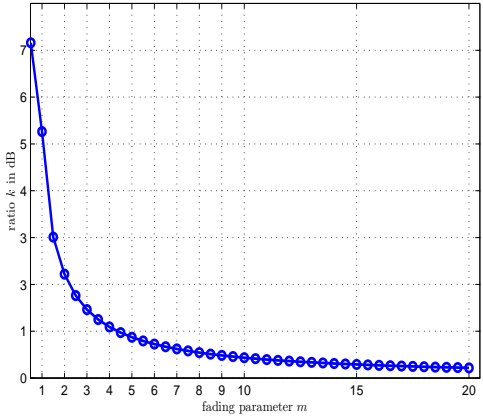}
	\caption{The ratio $k$ in (\ref{Eq:k}) versus the Nakagami $m$ parameter. When $m=1$, the SNR loss is about $5$dB. AS $m \rightarrow \infty$, $k \rightarrow 1$.}
	\label{fig:ratio}
\end{figure}

In Figure \ref{fig:ratio} we plot the loss due to fading when comparing with the AWGN case as a function of the Nakagami $m$ parameter in (\ref{Eq:k}). As expected from (\ref{Eq:comparison_1D}), when $m=1$, which means the fading is Rayleigh distributed, the SNR loss is about $5$dB. The loss decreases with increasing $m$ and converges to $1$.

\begin{figure}[t]
	\centering
	\includegraphics[
	height=2.4in,
	width=3in
	]
	{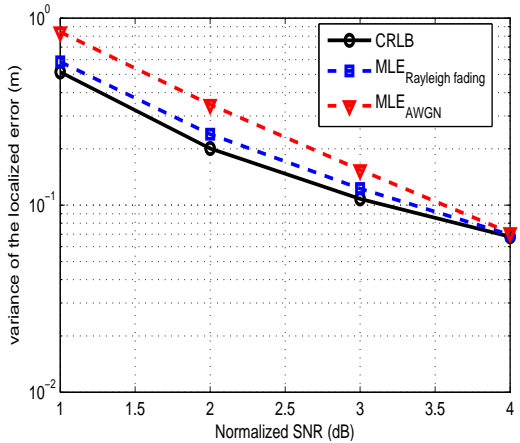}
	\caption{ML estimators comparison.}
	\label{fig:MLE}
\end{figure}

Finally, we compare estimators (\ref{Eq:ML_final}) and (\ref{Eq:ML_awgn}) both in the presence of fading by plotting the normalized SNR (with respect to $c^{2}$) vs. the variance of localization error in Figure \ref{fig:MLE}. We observe that the fading ML estimator (\ref{Eq:ML_final}) performs better than the AWGN ML estimator (\ref{Eq:ML_awgn}) in the presence of fading.

\subsection{Numerical Results for location detection}
\label{sec:simulations2}

\begin{figure}[t]
	\centering
	\includegraphics[
	height=2.4in,
	width=3in
	]
	{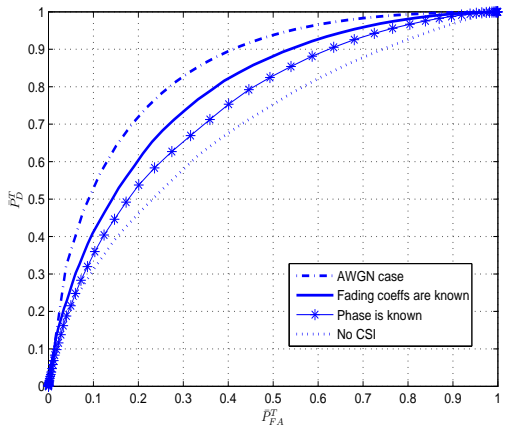}
	\caption{ROC curves for different scenarios.}
	\label{fig:det_comparisons}
\end{figure}

In the location detection formulation, we consider a $1\text{m} \times 1\text{m}$ square, $4$ anchors are at the corners, and $1$ node is in the middle of the square. The location of the node (when active) is known to all anchors. Figure \ref{fig:det_comparisons} shows the comparison between the AWGN case, fading coefficients are known to the anchors, the amplitude of fading coefficients are unknown but with a prior distribution, and the no CSI case. Here we fix the design parameter $K=1$ for all cases so that a single anchor's detection is sufficient for the FC to detect the node. One can see from the figure that the AWGN case outperforms all other cases as we expected, followed by the fading coefficients are known to anchors, the amplitude of fading coefficients are unknown but with a prior distribution, and no CSI case gives the worst performance. Intuitively speaking, we expect higher probability of detection when more information is available. 

\begin{figure}[t]
	\centering
	\includegraphics[
	height=2.4in,
	width=3in
	]
	{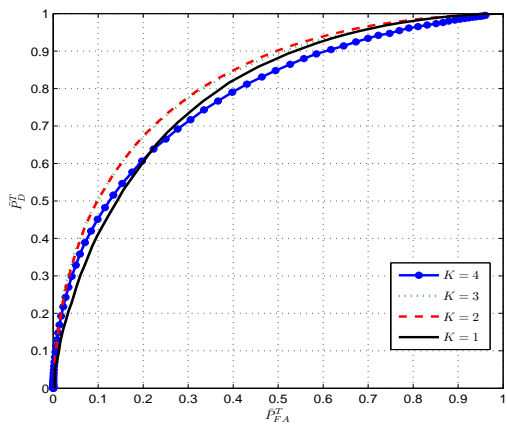}
	\caption{ROC curves when the fading coefficients are known to all anchors when $\text{SNR}=15\text{dB}$.}
	\label{fig:det_comparisons2}
\end{figure}

\begin{figure}[t]
	\centering
	\includegraphics[
	height=2.4in,
	width=3in
	]
	{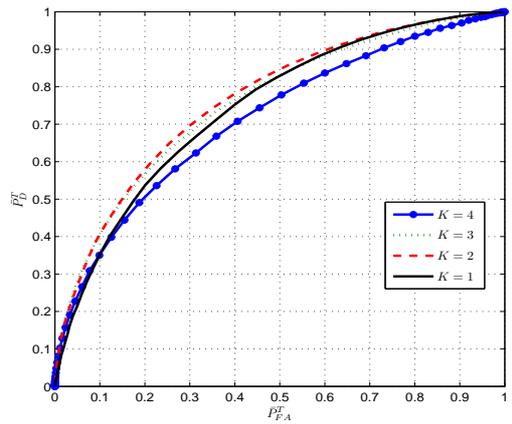}
	\caption{ROC curves when the amplitude of the fading coefficients are unknown to the anchors but with a prior distribution when $\text{SNR}=15\text{dB}$.}
	\label{fig:det_comparisons3}
\end{figure}

\begin{figure}[t]
	\centering
	\includegraphics[
	height=2.4in,
	width=3in
	]
	{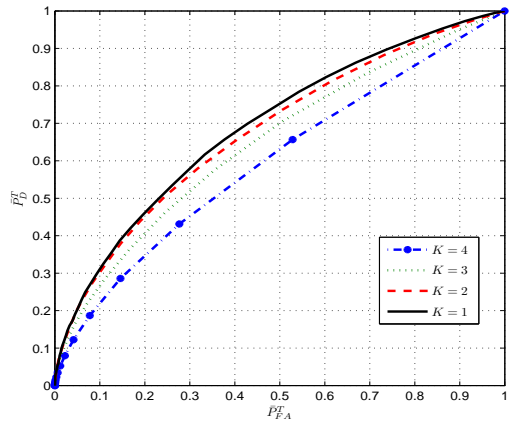}
	\caption{ROC curves when no CSI is available at any anchor when $\text{SNR}=15\text{dB}$.}
	\label{fig:det_comparisons4}
\end{figure}

Figure \ref{fig:det_comparisons2} through Figure \ref{fig:det_comparisons4} show the ROC curves for different cases as $K$ changes. From the figures one can see that for small $\Bpfa$, $K=4$ performs best, however as $\Bpfa$  increases, $K=4$ is not a good choice. Therefore, none of the $K$ values outperform others for all $\Bpfa$.

\begin{figure}[t]
	\centering
	\includegraphics[
	height=2.4in,
	width=3in
	]
	{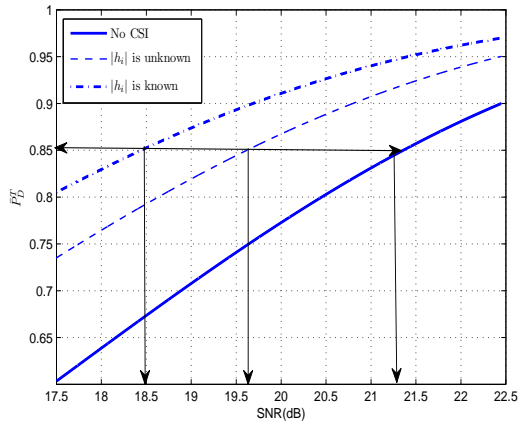}
	\caption{$\Bpd$ vs. SNR under different fading scenarios.}
	\label{fig:SNR}
\end{figure}

Figure \ref{fig:SNR} shows $\Bpd$ vs. SNR when $\Bpfa = 10^{-1}$ and $K=1$, and the amplitude of fading is Rayleigh distributed. From the figure one can see that to maintain the same $\Bpd=0.85$, the no CSI case needs $21.5$dB SNR, followed by the case when $\absa$ is unknown, which is about $19.6$dB, and $|h_{i}|$ is known needs the least amount of SNR, which is about $18.5$dB. Therefore, the SNR loss due to Rayleigh fading is about $2$dB, and unknown phase causes an additional $1$dB. However, one can also see that as $\Bpd$ increases, the SNR losses decrease.

\begin{figure}[t]
	\centering
	\includegraphics[
	height=2.4in,
	width=3in
	]
	{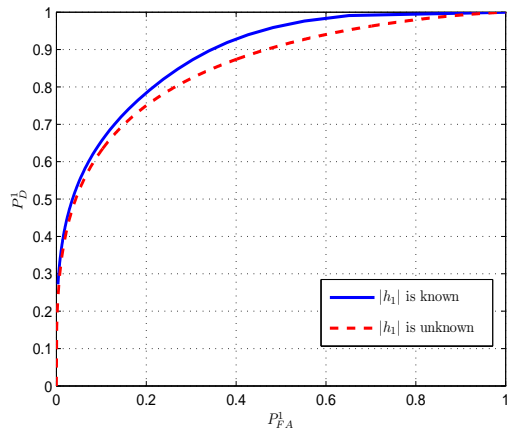}
	\caption{Comparisons between the threshold is a function of fading coefficients and the threshold is not a function of fading coefficients when $\text{SNR}=15\text{dB}$. Using the channel knowledge gives a better performance.}
	\label{fig:functionh}
\end{figure}

Figure \ref{fig:functionh} shows the ROC curve comparisons between (\ref{Eq:det_uncond_detector_highsnr}) and (\ref{Eq:det_Txa}) at the first anchor. From the figure we can see that by using the knowledge of the magnitude of the fading coefficients to set the threshold in (\ref{Eq:det_Txa}), the performance is better than the no CSI case (\ref{Eq:det_uncond_detector_highsnr}). 

\begin{figure}[t]
	\centering
	\includegraphics[
	height=2.4in,
	]
	{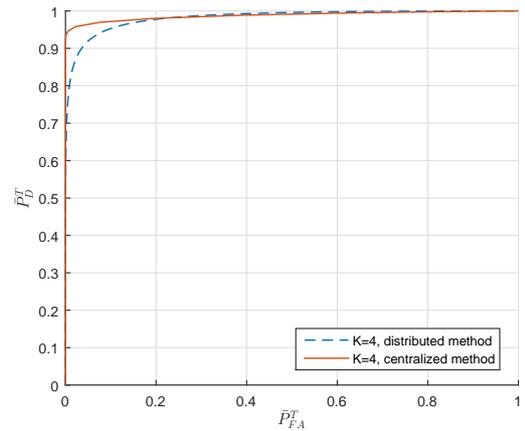}
	\caption{Comparison between the centralized detection scheme and distributed detection when the fading coefficients are known at each anchor. The design parameter is $K=4$. The penalty for using a distributed approach as compared with the centralized case is more pronounced at low $\bar{P}_{\text{FA}}$ values. This can be seen more clearly in Figure \ref{fig:pfa_zoom}. } 
	\label{fig:central}
\end{figure}

\begin{figure}[t]
	\centering
	\includegraphics[
	height=2.4in,
	]
	{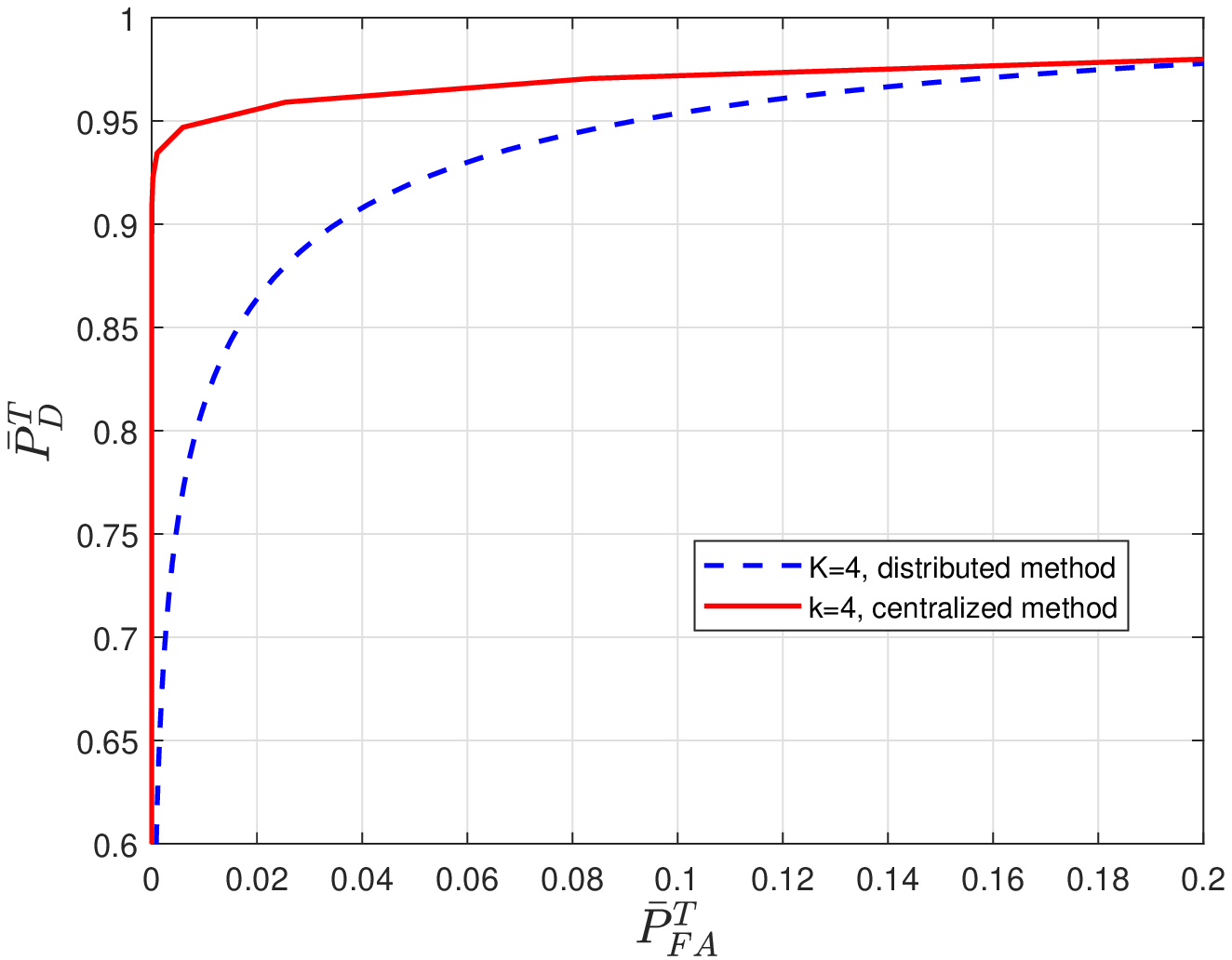}
	\caption{Comparison between the centralized detection scheme and distributed detection when the fading coefficients are known at each anchor. The design parameter is $K=4$. This shows in detail, the performance comparison seen in Figure \ref{fig:central}, at low $\bar{P}_{\text{FA}}$ values.}
	\label{fig:pfa_zoom}
\end{figure}

Figure \ref{fig:central} shows the comparison between the centralized detection scheme and the distributed detection scheme when the fading coefficients are known at each anchor case. We set the design parameter to be $K=4$. For the centralized detection scheme, each anchor transmits the measurements to a fusion center, the fusion center combines all the measurements and use the design parameter $K$ to make a final decision. From the figure, the penalty for using a distributed approach can be seen. The penalty for using a distributed approach as compared with the centralized case is more pronounced at low $\bar{P}_{\text{FA}}$ values. This can be seen more clearly in Figure \ref{fig:pfa_zoom}. 

\section{Conclusions}
\label{sec:conclusions}
In this paper, we considered both location estimation and location detection in wireless sensor networks (WSNs). To evaluate the performance of location estimation, the CRLB (Cramer-Rao lower bound) and the modified CRLB (MCRLB) are derived under different assumptions of channel knowledge. The results show that in both 1-D and 2-D WSNs, there is an SNR loss of about $5$dB compared to the AWGN case at high SNR. Under each assumption of channel knowledge for which the CRLB and MCRLB were derived, the ML estimator was derived. Results show that the ML estimator in the presence of fading has better performance than the ML estimator derived under the assumption of no fading, but used in a fading environment. 

In the detection formulation of the localization problem, each anchor makes its own decision on the presence of a signal from the target node, and transmits the decision to a fusion center. The fusion center needs at least $K$ anchors to agree that the node exists to detect the presence of the node, where $K$ is a design parameter. Three scenarios are considered: the fading coefficients are known at anchors; the phases of the fading coefficients are known but the amplitudes are unknown; and no CSI is available at any anchor. The ROC curves are plotted under different channel assumptions. From the plots we can see that the optimal $K$ depends on the requirements of $\Bpfa$ and $\Bpd$, and no particular $K$ value outperforms others for all $\Bpfa$. Finally, the simulation results show that using the knowledge of the fading coefficients to choose the threshold gives better performance.

\section{References}
\bibliographystyle{elsarticle-num}
\biboptions{sort&compress}
\bibliography{localization_bib}

\begin{thebibliography}{10}
\expandafter\ifx\csname url\endcsname\relax
  \def\url#1{\texttt{#1}}\fi
\expandafter\ifx\csname urlprefix\endcsname\relax\def\urlprefix{URL }\fi
\expandafter\ifx\csname href\endcsname\relax
  \def\href#1#2{#2} \def\path#1{#1}\fi

\bibitem{patwari}
N.~Patwari, J.~Ash, S.~Kyperountas, A.~Hero, R.~Moses, N.~Correal, Locating the
  nodes - cooperative location in wireless sensor networks, IEEE Signal
  Processing Magazine 22~(4) (2005) 54--69.

\bibitem{ville}
V.~Kaseva, M.~Kuorilehto, M.~Hannikainen, T.~Hamalainen, A wireless sensor
  network for {RF}-based indoor localization, in: EURASIP Journal on Advances
  in Signal Processing, 2008.

\bibitem{Han16}
G.~Han, J.~Jiang, C.~Zhang, T.~Duong, M.~Guizani, G.~Karagiannidis, A survey on
  mobile anchor node assisted localization in wireless sensor networks, 2016,
  pp. 1--25.

\bibitem{a2_unkerrors}
X.~Shi, G.~Mao, B.~D.~O. Anderson, Z.~Yang, J.~Chen, Robust localization using
  range measurements with unknown and bounded errors, IEEE Transactions on
  Wireless Communications 16~(6) (2017) 4065--4078.
\newblock \href {http://dx.doi.org/10.1109/TWC.2017.2691699}
  {\path{doi:10.1109/TWC.2017.2691699}}.

\bibitem{Ballardini}
A.~Ballardini, L.~Ferretti, S.~Fontana, A.~Furlan, D.~Sorrenti, An indoor
  localization system for telehomecare applications, 2015, pp. 1--11.

\bibitem{sensor}
F.~Zhao, L.~Guibas, Wireless sensor networks, an information processing
  approach, Morgan Kaufmann Publishers, 2004.

\bibitem{Spanias}
A.~Spanias, Digital signal processing; an interactive approach - 2nd edition,
  ISBN: 978-1-4675-9892-7, Lulu press ondemand publishers, 2014.

\bibitem{intro_positioning}
D.~Niculescu, Positioning in ad hoc sensor networks, in: Network, IEEE, 2004,
  pp. 24--29.

\bibitem{Marc}
M.~Willerton, M.~Banavar, X.~Zhang, A.~Manikas, C.~Tepedelenlioglu, A.~Spanias,
  T.~Thomoton, E.~Yeatman, C.~A., Sequential wireless sensor network discovery
  using wide aperture array signal processing, in: European Signal Processing
  Conference, Bucharest, 2012, pp. 2278--2282.

\bibitem{Poor}
S.~Gezici, H.~Poor, Position estimation via ultra-wideband signals, in: IEEE
  Special issues on UWB Technology and Emerging Applications, 2008.

\bibitem{Mao}
G.~Mao, B.~Fidan, Localization Algorithms and Strategies for Wireless Sensor
  Networks, Information Science Reference, 2009.

\bibitem{Neal03}
N.~Patwari, A.~O. {Hero III}, M.~Perkins, N.~Correal, R.~O'Dea, Relative
  location estimation in wireless sensor networks, IEEE Transactions on Signal
  Processing 51~(8) (2003) 2137--2148.

\bibitem{RSS}
X.~Li, {RSS}-based location estimation with unknown pathloss model, IEEE
  Transactions on Wireless Communications (2006) 3626--3633.

\bibitem{ref3_Liang}
C.~Liang, F.~Wen, Received signal strength-based robust cooperative
  localization with dynamic path loss model, IEEE Sensors Journal 16~(5) (2016)
  1265--1270.

\bibitem{ref1_Cota}
J.~Cota-Ruiz, J.~Rosiles, P.~Rivas-Perea, E.~Sifuentes, A distributed
  localization algorithm for wireless sensor networks based on the solutions of
  spatially-constrained local problems, IEEE Sensors Journal 13~(6) (2013)
  2181--2191.

\bibitem{ref2_Wang}
B.~Wang, G.~Wu, S.~Wang, L.~Yang, Localization based on adaptive regulated
  neighborhood distance for wireless sensor networks with a general radio
  propagation model, IEEE Sensors Journal 14~(11) (2014) 3754--3762.

\bibitem{Xue3}
X.~Zhang, M.~Banavar, M.~Willerton, A.~Manikas, C.~Tepedelenlio\u{g}lu,
  A.~Spanias, T.~Thornton, E.~Yeatman, A.~Constantinides, Performance
  comparison of localization thechniques for sequential {WSN} discovery, in:
  Sensor Signal Processing for Defence (SSPD), London, 2012, pp. 1--5.

\bibitem{a6_rss_overview}
R.~Niu, A.~Vempaty, P.~K. Varshney, Received-signal-strength-based localization
  in wireless sensor networks, Proceedings of the IEEE 106~(7) (2018)
  1166--1182.
\newblock \href {http://dx.doi.org/10.1109/JPROC.2018.2828858}
  {\path{doi:10.1109/JPROC.2018.2828858}}.

\bibitem{a3_rlsnlos}
G.~Wang, A.~M. So, Y.~Li, Robust convex approximation methods for tdoa-based
  localization under nlos conditions, IEEE Transactions on Signal Processing
  64~(13) (2016) 3281--3296.
\newblock \href {http://dx.doi.org/10.1109/TSP.2016.2539139}
  {\path{doi:10.1109/TSP.2016.2539139}}.

\bibitem{FoutzDOA}
J.~Foutz, A.~Spanias, M.~K. Banavar, Narrowband Direction of Arrival Estimation
  for Antenna Arrays, Morgan and Claypool, 2008.

\bibitem{ManikasLAA}
A.~Manikas, Y.~I. Kamil, M.~Willerton, Source localization using sparse large
  aperture arrays, IEEE Transactions on Signal Processing 60~(12) (2012)
  6617--6629.
\newblock \href {http://dx.doi.org/10.1109/TSP.2012.2210886}
  {\path{doi:10.1109/TSP.2012.2210886}}.

\bibitem{a1_RSS_AoA}
S.~Tomic, M.~Beko, R.~Dinis, Distributed rss-aoa based localization with
  unknown transmit powers, IEEE Wireless Communications Letters 5~(4) (2016)
  392--395.
\newblock \href {http://dx.doi.org/10.1109/LWC.2016.2567394}
  {\path{doi:10.1109/LWC.2016.2567394}}.

\bibitem{patent}
X.~Zhang, C.~Tepedelenlio\u{g}lu, M.~Banavar, A.~Spanias, Maximum likelihood
  localization in the presence of channel uncertainties, ISBN: 9781598297010,
  {Predisclosure AzTE}, 2013.

\bibitem{Bergamo}
P.~Bergamo, G.~Mazzini, Localization in sensor networks with fading and
  mobility, in: IEEE International Symposium on Personal, Indoor and Mobile
  Radio Communications, 2002.

\bibitem{fading}
S.~Sattarzadeh, B.~Abolhassani, {TOA} extraction in multipath fading channels
  for location estimation, in: IEEE International Symposium on Personal,
  Indoor, and Mobile Radio Communications Conference, 2006.

\bibitem{Van1}
N.~Vankayalapati, S.~Kay, D.~Quan, {TDOA} based direct positioning maximum
  likelihood estimator and the cramer-rao bound, IEEE Transactions on Aerospace
  and Electronic Systems 50~(3) (2014) 1616--1635.

\bibitem{a4_slat}
B.~Zhou, Q.~Chen, P.~Xiao, The error propagation analysis of the received
  signal strength-based simultaneous localization and tracking in wireless
  sensor networks, IEEE Transactions on Information Theory 63~(6) (2017)
  3983--4007.
\newblock \href {http://dx.doi.org/10.1109/TIT.2017.2693180}
  {\path{doi:10.1109/TIT.2017.2693180}}.

\bibitem{a5_lbmse}
I.~Bergel, Y.~Noam, Lower bound on the localization error in infinite networks
  with random sensor locations, IEEE Transactions on Signal Processing 66~(5)
  (2018) 1228--1241.

\bibitem{Ray06}
S.~Ray, W.~Lai, I.~Paschalidis, Statistical location detection with sensor
  networks, IEEE Transactions on Information Theory 52~(6) (2006) 2670--2683.

\bibitem{Van2}
N.~Vankayalapati, S.~Kay, Asymptotically optimal detection of low probability
  of intercept signals using distributed sensors, IEEE Transactions on
  Aerospace and Electronic Systems 48~(1) (2012) 737--748.

\bibitem{xue_det}
X.~Zhang, C.~Tepedelenlioglu, M.~Banavar, A.~Spanias, Distributed location
  detection in wireless sensor networks, in: Asilomar Conference on Signals,
  Systems and Computers, 2013, pp. 428--432.

\bibitem{RSSdetection}
S.~Yan, R.~Malaney, I.~Nevat, G.~Peters, Signal strength based location
  verification under spatially correlated shadowing, in: IEEE International
  Conference on Communications, 2014, pp. 2617--2623.

\bibitem{vantrees}
H.~{Van Trees}, Detection, Estimation and Modulation Theory, John Wiley and
  Sons, Inc., 1968.

\bibitem{Kay}
S.~Kay, Fundamentals of Statistical Signal Processing - {Volume} {II} Detection
  Theory, Printice Hall, 1998.

\bibitem{Helstrom}
C.~Helstrom, Elements of Signal Detection and Estimation, Prentice Hall, 1994.

\bibitem{MCRLB}
A.~Andrea, U.~Mengali, R.~Reggiannini, The modified cramer-rao bound and its
  application to synchronization problems, IEEE Transactions on Communications
  42~(234) (1994) 1391--1399.

\bibitem{wireless}
A.~Goldsmith, Wireless {Communications}, Combridge University Press, 2005.

\bibitem{tables}
I.~Gradshteyn, I.~Ryzhik, Tables of Integrals, Series, and Products, Elsevier
  Inc., 2007.

\bibitem{Zekavat11}
R.~Zekavat, M.~Buehrer, Handbook of Position Location: Theory, Practice and
  Advances, Wiley-IEEE Press, 2011.

\bibitem{bandpass}
A.~Ong, Bandpass analog-to-digital conversion for wireless applications, Tech.
  rep., Stanford University (1998).

\end{thebibliography}

\appendix

\section{ Proof of Theorem I}
\label{sec:appx}
To prove the Theorem, we express the Lagrangian as
\begin{equation}
\begin{split}
\label{Eq:lag1}
\int^{\infty}_{0}Q\left(\gammaprime\left(\absa\right)-\sqrt{\frac{\absaa}{\Sigma}}\right)f_{|h_{i}|^{2}}\left(x\right)dx + \\
\lambda\left(\int^{\infty}_{0}Q\left(\gammaprime\left(\absa\right)\right)f_{|h_{i}|^{2}}\left(x\right)dx-\alpha\right).
\end{split}
\end{equation}
Taking the derivative of (\ref{Eq:lag1}) with respect to $\gammaprime$, we have
\begin{equation}
\label{Eq:proof2}
\begin{split}
\int^{\infty}_{0} \frac{\partial Q\left(\gammaprime\left(\absa\right)-\sqrt{\frac{\absaa}{\Sigma}}\right)}{\partial \gammaprime\left(\absa\right)} f_{|h_{i}|^{2}}\left(x\right)dx +  \\
\lambda \int^{\infty}_{0} \frac{\partial Q\left(\gammaprime\left(\absa\right)\right)}{\partial \gammaprime\left(\absa\right)} f_{|h_{i}|^{2}}\left(x\right)dx.
\end{split}
\end{equation}
Setting (\ref{Eq:proof2}) to $0$, applying the formula $\frac{d Q(x)}{dx}=-\frac{1}{\sqrt{2\pi}}\exp\left(-\frac{x^{2}}{2}\right)$, and solving for $\gammaprime$, we have
\begin{equation}
\label{Eq:lag2}
\gammaprime\left(\absa\right)=\ln \lambda \sqrt{\frac{\Sigma}{\absaa}}+\frac{1}{2}\sqrt{\frac{\absaa}{\Sigma}}.
\end{equation}
Setting $\lambda=\gamma$, (\ref{Eq:lag2}) is equivalent to (\ref{Eq:det_gammaprime}), which proves Theorem I.

\end{document}